\documentclass[aps,prb,twocolumn,a4paper,floatfix,showpacs]{revtex4-1}
\usepackage[version=3]{mhchem}

\usepackage{times,mathptmx}
\usepackage{graphicx} 
\usepackage{float}
\usepackage{array}
\usepackage[T1]{fontenc}
\usepackage[usenames,dvipsnames]{xcolor}
\usepackage{hyperref}
\usepackage{epstopdf}

\usepackage{threeparttable}
\usepackage{physics}
\usepackage{siunitx}
\usepackage{multirow}
\usepackage{booktabs} 

\newcommand*{\olcite}[1]{\onlinecite{#1}} %

\newcommand*{\veck}{{\mathbf{k}}}

\newcommand*{\RMT}{R_{\rm MT}}
\newcommand*{\nLO}{n_{\text{LO}}}
\newcommand*{\dlLO}{\Delta l_{\rm LO}}

\newcommand*{\lmlo}{l^{(\text{LO})}_{\rm max}}
\newcommand*{\lmv}{l^{(\text{v})}_{\rm max}}

\begin{document}

\title{Electronic Band Structure of Cuprous and Silver Halides: a Numerically Accurate All-Electron $GW$ Study}
\author{Min-Ye Zhang}
\author{Hong Jiang}
\thanks{To whom any correspondences should be addressed (Email: jianghchem@pku.edu.cn).}
\affiliation{Beijing National Laboratory for Molecular Sciences, College of Chemistry and Molecular Engineering, Peking University, 100871 Beijing, China}
\date{\today}

\begin{abstract} 
Group IB metal halides (CuX and AgX, X=Cl, Br and I) are widely used in optoelectronic devices and photochemical catalysis due to their appropriate optical and electronic properties. First-principles calculations have confronted difficulties in accurately predicting their electronic band structures. Here we study CuX and AgX with many-body perturbation theory in the $GW$ approximation, implemented in the full-potential linearized augmented plane waves (FP-LAPW) framework. Comparing the quasi-particle band structures calculated with the default LAPW basis and the one extended by high-energy local orbitals (HLOs), denoted as LAPW+HLOs, we find that it is crucial to include HLOs to achieve sufficient numerical accuracy in $GW$ calculations of these materials. Using LAPW+HLOs in semi-local density functional approximation based $GW_0$ calculations leads to good agreement between theory and experiment for both band gaps and the splitting between metal (Cu or Ag) $d$ and X-$p$ states. It is indicated that quasi-particle band structures of CuX and AgX are crucially influenced by the numerical accuracy of $GW$ implementations, similar to what was found in ZnO [Jiang, H.; Blaha, P. \textit{Phys. Rev. B} 2016, \textbf{93}, 115203]. This work emphasizes the importance of numerical accuracy in the description of unoccupied states for quasi-particle band structure of materials with the $d^{10}$ electronic configuration.
\end{abstract}

\maketitle

\section{Introduction}

Cuprous and silver halides (CuX and AgX, X=Cl, Br, I) have been redrawing increasing practical interests during the past decades for their interesting optical and electronic properties.
Cuprous halides are wide-gap semiconductors with large exciton binding energy, and are promising candidates for applications in optoelectronic devices.\cite{Valenta2001, Ahn2013, Ahn2016, Azhikodan2017}
In particular, being a native p-type semiconductor,\cite{Wang2011} the transparent CuI film has not only been employed as a hole transport material in solar cells,\cite{Grundmann2013, Christians2014, Sepalage2015} but also shown exceptional performance as a thermoelectric material.\cite{Yang2017} Silver halides has been used in light conversion since mid-1800s, owing to their high photosensitivity. They are the first photographic materials \cite{Tani1995} and constitute the first photovoltaic solar cell designed by E. Becquerel.\cite{Becquerel1839, Williams1960} Recently, silver halides have been extensively exploited in various scenarios of photocatalysis,\cite{An2016} such as \ce{CO2} reduction,\cite{An2012} degradation of organic pollutants\cite{Cai2013, Zhang2015} and water splitting.\cite{Lou2012} However, despite their wide applications, a thorough theoretical understanding of fundamental properties of this class of materials is still lacking, e.g. the phase transition of CuI at high temperature,\cite{Hull1994, Zhu2012} the extraordinarily large excitonic binding energy of CuX,\cite{Goldmann1977, Azhikodan2017} and the electronic dynamics within AgX in the latent image formation.\cite{Marchetti1992, Tani2007, Loftager2016}

Nowadays, first-principles electronic structure calculations are being practiced routinely to predict electronic and optical properties of materials. Among different methods, Kohn-Sham (KS) density functional theory (DFT) in the local density approximation or generalized gradient approximation (LDA/GGA) is most widely used for its efficiency and accuracy. However, stemming from the self-interaction error (SIE) in the LDA/GGA, the band gaps of semiconductors are systematically underestimated or even predicted to be negative, i.e. qualitatively wrong metallic state, which deteriorates the reliability of the predictions by practical LDA/GGA based DFT calculations. Previous work confirmed that the band gaps predicted for cuprous and silver halides by LDA/GGA are typically smaller than experimental values by 1-2 eV,\cite{Victora1997, Vogel1998, Wilson2008, Azhikodan2017} and the problem is only partly remedied when using the hybrid functional approximation.\cite{Loftager2016, Pishtshev2017}

The many-body perturbation theory based on Green's function in the $GW$ approximation has proven to be able to accurately predict electronic band structure of typical semiconductors, \cite{Hybertsen1986, Godby1988, Aryasetiawan1998} and it has been applied in attempt to resolve the band-gap problem in cuprous and silver halides.\cite{vanSchilfgaarde2006, Pishtshev2017, Azhikodan2017, vanSetten2017, Gao2018}
However, LDA/GGA-based $G_0W_0$ calculations generally give underestimated band gaps for these materials.\cite{vanSchilfgaarde2006, Azhikodan2017, vanSetten2017, Gao2018}
Particularly in CuX, the error ranges from 0.7 to 2.7 eV for the band gap at $\Gamma$,\cite{vanSchilfgaarde2006, vanSetten2017, Gao2018} with the largest error observed in CuCl.\cite{vanSetten2017}
Although it is well known that one-shot $GW$ calculations based on LDA/GGA tend to underestimate the band gaps for semiconductors,\cite{Faleev2004, Shishkin2007-PRB} it is inferred by the exceptionally large error that some essential ingredients may be missing in the employed LDA/GGA-based $G_0W_0$ implementation to predict accurate band gaps for the cuprous compounds. Previous $GW$ results will be discussed later in more details along with those obtained in the present work.

It is worth noticing that considerably underestimated band gap predicted by full-frequency one-shot $GW$ calculation has been observed as well in the wide-band-gap semiconductor zinc oxide (ZnO) with shallow $d$-states, and has raised a continuing debate on the validity of the approximation and implementation adopted.\cite{Shih2010, Friedrich2011a, Friedrich2011b, Stankovski2011, Miglio2012, Jiang2016, Chu2016, Nabok2016, Zhang2016, Cao2017} Within the framework of all-electron $GW$ calculations based on linearized augmented-plane-wave (LAPW) basis,\cite{Andersen1975} it has been shown that the culprit for the problem is the inadequate description of high-lying states to be summed over due to the linearization error.\cite{Friedrich2011} Recently, Jiang and Blaha found that by extending the normally used LAPW basis with additional high energy local orbitals (HLOs) of energy up to a few hundred Rydberg above the Fermi level and large angular quantum numbers (with $l$ up to 6 or larger), one can obtain $GW$  quasi-particle (QP) band gap of ZnO in close agreement with experiment even at the LDA/GGA-based $G_0W_0$ or $GW_0$ level without sacrificing the accuracy for other ``simpler'' sp semiconductors.\cite{Jiang2016} When using the HLOs-extended LAPW basis, the $GW_0$ approach using the LDA/GGA plus the Hubbard $U$ correction (DFT+$U$) as the reference can also describe electronic band structure of strongly correlated $d$- or $f$-electron oxides very well.\cite{Jiang2018} It is therefore natural to consider whether the inclusion of HLOs in $GW$ calculations can also solve the band gap problem of cuprous and silver halides.

In this work, we present the all-electron $GW$ calculations in the LAPW framework for cuprous and silver halides. We compare the results obtained from using the standard LAPW basis and those from using HLOs-extended LAPW basis, and carefully analyze the effects of including HLOs on various aspects of electronic band structure of these materials. The rest part of the paper is organized as follows. The computational details of the all-electron $GW$ calculations are given in the next section. Then we present our main results on quasi-particle band structure of cuprous and silver halides and compare them with available experiment data in Sec.III. Sec. VI summarize our main findings. 

\section{Computational method and details}\label{s:method-and-details}

\subsection{Crystal structures of CuX and AgX}
To make the comparison between the calculated results with the data extracted from low-temperature experiments meaningful, we use the thermodynamically stable crystal structures with the experimental lattice constants whenever available. The crystal phases and corresponding lattice constants of the cuprous and silver halides used in our calculations are summarized in Table \ref{tab:structs}. It should be mentioned that at low temperature, zincblende AgI ($\gamma$-AgI) is metastable and forms mixture with the wurtzite phase ($\beta$-AgI). Nevertheless, we focus on the zincblende phase.
\begin{table}
    \centering
    \caption{\ Structure and lattice constants of cuprous and silver halides used in the study. ``ZB'' and ``RS'' stands for the structure of zincblende and rocksalt, respectively. \label{tab:structs} }
    \begin{tabular}{lccc}
        \hline \hline
        Systems & Type & Lattice constants (\AA) & Ref. \\
        \hline
        CuCl & ZB & 5.420 & \olcite{Hull1994} \\ 
        CuBr & ZB & 5.677 & \olcite{Altorfer1994} \\
        CuI  & ZB & 6.052 & \olcite{Keen1995} \\
        AgCl & RS & 5.550 & \olcite{Berry1955} \\
        AgBr & RS & 5.775 & \olcite{Berry1955} \\ 
        AgI  & ZB & 6.499 & \olcite{Keen1993} \\ 
      \hline \hline
    \end{tabular}
\end{table}

\subsection{$GW$ method with LAPW basis extended by HLOs}
We use the all-electron $GW$ method implemented in the HLOs-extended LAPW basis to calculate the quasi-particle band structures of CuX and AgX. The basic theory and detailed formalism employed in the implementation has been presented in our previous work.\cite{Jiang2013a, Jiang2016} The HLOs are generated following the way described by Laskowski and Blaha.\cite{Laskowski2012} The inclusion of HLOs has been demonstrated to produce significantly more accurate quasi-particle band structures for typical $sp$ semiconductors,\cite{Jiang2016} later transition metal mono-oxides and $f$-electron oxides,\cite{Jiang2018} compared to the results obtained from using the standard LAPW basis. The improvement can be attributed to a more accurate and complete consideration of unoccupied states in the high-energy regime. The inaccuracy of high-lying unoccupied states is due to the linearization error of the LAPW basis functions, which presents no essential obstacles for DFT calculations with LDA/GGA or hybrid functionals, since only occupied and low-lying unoccupied states are used and they are accurately described by the standard LAPW basis. However, for $GW$ and DFT with rung-5 density functional approximations,\cite{SuNQ2017} such as the random phase approximation (RPA) for the ground state total energy,\cite{Ren2012b} which involve the summation of unoccupied states, the completeness of the summation and the quality of these states play a crucial role in the numerical accuracy.\cite{Grueneis2014, Cui2016} Both factors are taken into account by including additional local orbitals energetically much higher than the Fermi level to the standard LAPW basis. We term this extended basis as LAPW+HLOs.\cite{Jiang2016}

The quality of LAPW+HLOs is controlled by two parameters, besides those of the standard LAPW basis, namely the additional number of nodes in the radial function of highest energy local orbitals with respect to that of the LAPW basis with the same angular quantum number, denoted as $\nLO$, and the maximum angular quantum number of local orbitals, denoted as $\lmlo$. In general, the larger $\nLO$ and $\lmlo$, the higher the HLOs reach in the energy space. From a real space point of view, $\nLO$ and $\lmlo$ characterize the radial and angular variation of local orbitals within the muffin-tin sphere, respectively. We denote the default LAPW basis by $\nLO=0$ in the recent version of \textsc{WIEN2k},\cite{wien2k-2001} which is actually a mixture of the APW+lo basis\cite{Madsen2001} for the valence states, the ordinary LAPW basis for higher $l$ channels up to $l_\text{max}=10$ and additional local orbitals (LOs) for semi-core states if present.\cite{wien2k-2001} By default, we add HLOs to the LAPW basis with the angular momentum $l$ up to $\lmlo = \mathrm{min}\{3,\lmv+1\}$, with $\lmv$ being the largest $l$ of valence orbitals for each element, e.g. $\lmv=1$ for Cl and 2 for the other elements, i.e. Br, I, Cu and Ag.\cite{Laskowski2012, Laskowski2014} The convergence with respect to both $\nLO$ and $\lmlo$, the latter being represented by $\dlLO$ in $\lmlo \equiv \lmv + \dlLO$, is investigated. The convergence test is performed with $\Gamma$-centered Monkhorst-Pack $\veck$-mesh of $2\times2\times2$. 

$GW$ results in both $G_0W_0$ and $GW_0$ schemes are presented, where Kohn-Sham orbital energies and wave functions calculated with the Perdew-Burke-Ernzerhof (PBE) \cite{Perdew1996} GGA are used as the input to calculate one-body Green's function $G$ and screened Coulomb interaction $W$. All available empty states are used in the summation of states for the calculation of screened interaction and self-energy. For the sampling of the Brillouin zone, a $6\times6\times6$ $\Gamma$-centered $\veck$-mesh is employed for $GW$ calculations with the standard LAPW basis. Considering that $GW$ calculations with LAPW+HLOs are expensive at a dense $\veck$-mesh, and to reduce the computational cost without sacrificing numerical accuracy, the quasi-particle band gaps with LAPW+HLOs on the fine $\veck$-mesh (here $6^3$) is obtained by shifting the gap calculated from the default LAPW basis by the correction in a coarser $\veck$-mesh (here $4^3$) according to
\begin{equation}\label{eq:gap-shift-for-HLOs}
\begin{aligned}
E^{GW,\text{HLOs}}_{\text{g}}(6^3) = E^{GW}_{\text{g}}(6^3) + \left[E^{GW,\text{HLOs}}_{\text{g}}(4^3) - E^{GW}_{\text{g}}(4^3)\right],
\end{aligned}
\end{equation}
The quasi-particle band structure diagram along a particular $\veck$-path is obtained by interpolating the quasi-particle energy levels calculated with the $4\times4\times4$ $\Gamma$-centered Monkhorst-Pack $\veck$-mesh using the Fourier interpolation technique.\cite{Pickett1988}

The present all-electron $GW$ calculations are performed by the $GW$ facilities in the \textsc{GAP2} program,\cite{Jiang2013a, Jiang2016} interfaced to \textsc{WIEN2k}\cite{wien2k-2001}.

\subsection{Density Functional Calculations for Band Structure}
For comparison, DFT calculations with PBE\cite{Perdew1996} semi-local approximation and the HSE06\cite{Heyd2003, Heyd2006} hybrid functional approach are performed by \textsc{WIEN2k}.\cite{wien2k-2001}
The energies and wave functions of Kohn-Sham orbitals from PBE are also used as starting point for $GW$ computation. Hybrid functional calculations are performed by using the second-variational procedure.\cite{Tran2011}

For self-consistent-field (SCF) calculations, a $10\times10\times10$ $\Gamma$-centered $\veck$-mesh is employed for numerical integration over the first Brillouin zone of the primitive cell of the face-centered cubic crystal, corresponding to 47 points in the irreducible Brillouin zone (IBZ) of both rocksalt and zincblende structures. The criterion for energy convergence is set to $10^{-6}$ Rydberg (Ry). For the basis expansion, $R_{\text{MT,min}}K_{\text{max}}=7.0$ is chosen for the plane wave cutoff in the interstitial region, where $R_{\text{MT,min}}$ is the minimal muffin-tin radius $\RMT$. In the present study, 2.1 and 2.3 Bohr are chosen as $\RMT$ for non-iodine elements and iodine, respectively. The default LAPW basis (i.e. $\nLO=0$) is used at this stage, since the effects of including HLOs in SCF calculations are negligible, as we have shown in a previous study.\cite{Jiang2016} A similar interpolation technique as described previously is employed to obtain the band structure along a particular $\veck$-path for comparison with $GW$. Considering that the systems investigated in this work are composed of heavy elements, we also consider the effects of spin-orbit coupling (SOC) on electronic band structure by using the second variational approach \cite{Singh2006} at the PBE level.

\section{Results and Discussion}\label{s:result-and-dissucssion}

\subsection{Importance of including HLOs}
\begin{figure}[ht]
    \centering
    \includegraphics[width=0.8\linewidth]{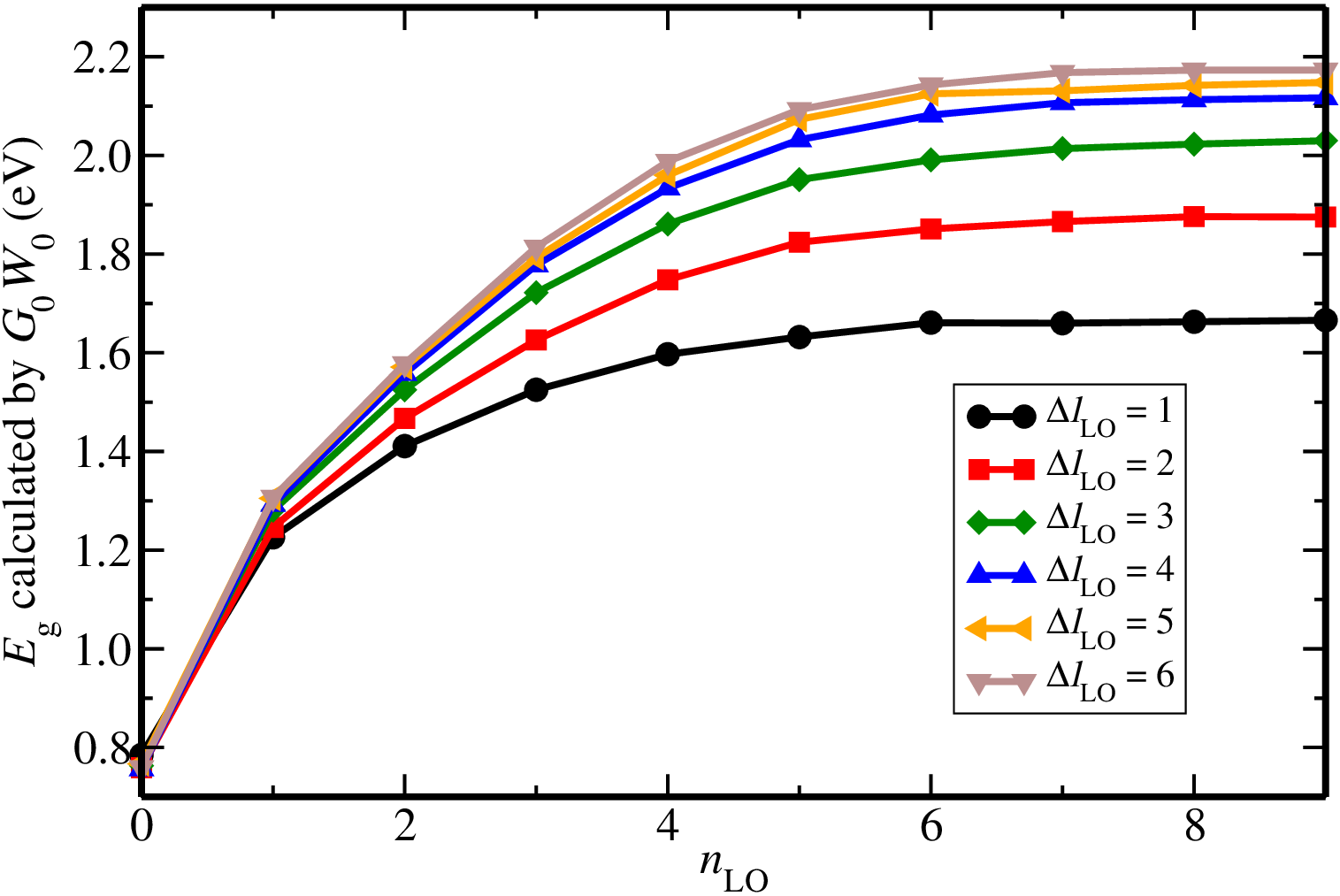}
    \caption{\ Fundamental band gap of CuCl calculated by $G_0W_0$@PBE with different HLOs setups, characterized by $\nLO$ and $\dlLO$.}
    \label{fig:CuCl-g0w0-gap-vs-hlo}
\end{figure}

We first discuss the convergence of $GW$ band gaps with respect to the setting of HLOs, namely $\nLO$ and $\dlLO$ (see the previous section for the definition). As the $GW$ calculation with many HLOs is computationally demanding, it is preferable to use minimal HLOs to achieve the required accuracy. Since the effects of including HLOs on the $GW$ results are system-dependent and a detailed guide for such setup is not available currently, the convergence issue of all the systems considered in this work have been investigated to obtain some insights.
We present the results of CuCl as an example here, and those of other materials considered can be found in the supplemental material.\footnote{See Supplemental Material at [URL] for results of HLO convergence tests for CuBr, CuI and AgX.}

To begin with, we investigate how the fundamental band gap (direct at the $\Gamma$ point) predicted by $G_0W_0$ within LAPW+HLOs changes with $\nLO$ and $\dlLO$. As shown in Fig. \ref{fig:CuCl-g0w0-gap-vs-hlo}, the gap increases significantly as either $\nLO$ or $\dlLO$ increases. Moreover, the speed of convergence with respect to one parameter is strongly dependent on the value of the other. The band gap increases by 0.51 eV when $\dlLO$ is changed from 1 to 6 at $\nLO=8$, which is about 6 times larger than that at $\nLO=1$ (0.08 eV). Considering the convergence with respect to $\nLO$, the band gap changes by 0.86 eV when $\nLO$ increases from 1 to 8 at $\dlLO=6$, which is 2 times larger than that at $\dlLO=1$ (0.43 eV). The $G_0W_0$ band gap of CuCl is converged within 0.05 eV for $\nLO=6, \dlLO=5$, in a sense that the change is smaller than 0.05 eV when further increasing both $\nLO$ and $\dlLO$ by 1.

\begin{figure}[!htp]
    \centering
    \includegraphics[width=0.8\linewidth]{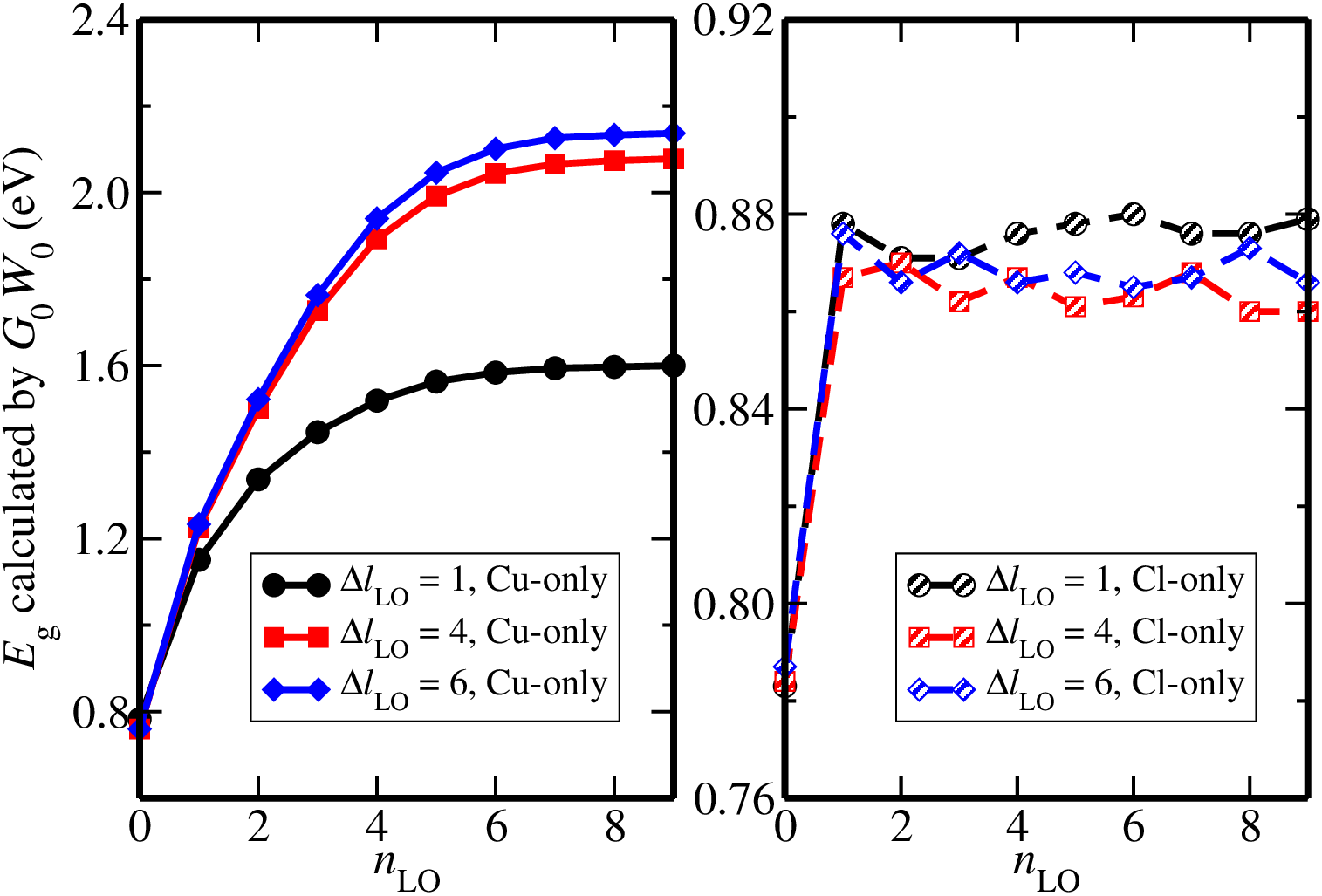}
    \caption{\ Fundamental band gap of CuCl calculated by $G_0W_0$@PBE with different HLOs setups on either Cu or Cl atom.}
    \label{fig:elem-dependence-CuCl-g0w0-gap}
\end{figure}

The above discussions are based on the results obtained with HLOs added to both Cu and Cl atoms. In a previous study, we have shown in ZnO and ZnS that the effects on $GW$ band gap of including HLOs depends on the element to which HLOs are added and that the effects on different elements are additive to some extent, i.e. the summation of the changes in the band gap with HLOs added to each element separately is nearly equal to the change with HLOs added to all elements simultaneously.\cite{Jiang2016} According to this observation, we perform the calculations with HLOs added only to either Cu or Cl atom, and the results are shown in Fig. \ref{fig:elem-dependence-CuCl-g0w0-gap}. It can be seen that the $GW$ band gap is very sensitive to the HLOs on Cu atom and the convergence behavior with respect to $\nLO$ and $\dlLO$ is very similar to that when HLOs are added to both atoms. On the other hand, when HLOs are set on Cl, the $G_0W_0$ gap increases by 0.10 eV when the HLOs setting changes from $\nLO=0$ (the default LAPW basis) to $\nLO=1, \dlLO=1$, and remains essentially unchanged when further increasing $\nLO$ or $\dlLO$. We can then infer that in order to obtain numerically accurate $G_0W_0$ gap of CuCl, it is not necessary to add HLOs with $\nLO$ and $\dlLO$ on Cl as large as those of HLOs on Cu. Similar conclusions can be drawn for the other materials. To balance the computational workload and numerical accuracy, we choose HLOs with $\nLO=8,\dlLO=6$ for Cu and $\nLO=2,\dlLO=4$ for X in CuX, and those with $\nLO=8,\dlLO=5$ for Ag and $\nLO=2,\dlLO=4$ for X in AgX, which can achieve 0.05 eV convergence for the $G_0W_0$ or $GW_0$ band gaps of all systems considered in this work. Unless stated otherwise, the notation LAPW+HLOs for any practiced calculations refers to this HLO setup in the remaining part of the article.

\begin{figure}[!htp]
    \centering
    \includegraphics[width=0.8\linewidth]{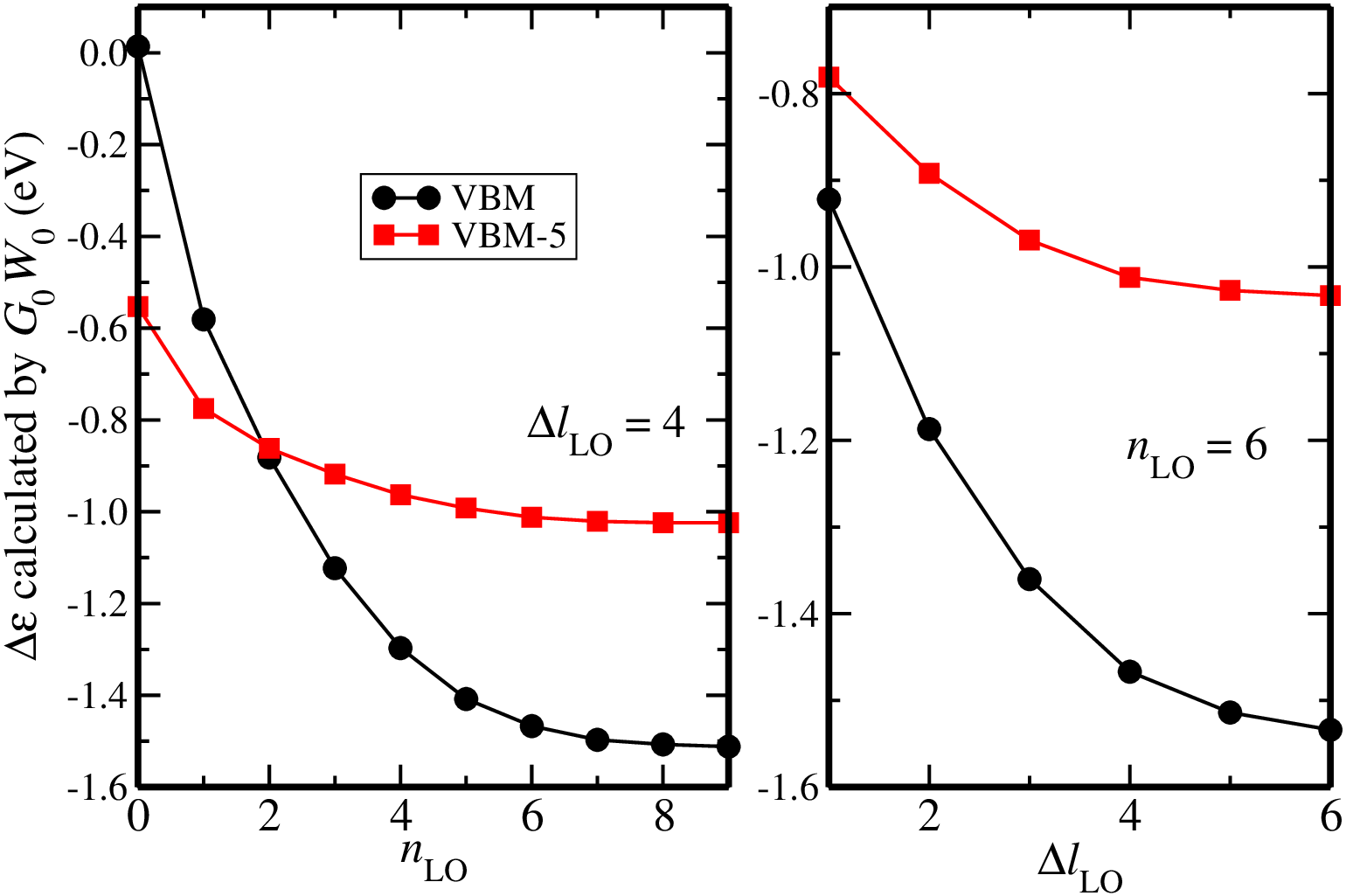}
    \caption{\ Dependence of $G_0W_0$ QP correction to the valence states at the $\Gamma$ point on $\nLO$ and $\dlLO$ of HLOs setup in CuCl. $\dlLO$ and $\nLO$ are fixed as 4 and 6, respectively, when the other parameter is varying.}
    \label{fig:CuCl-g0w0-degw-hlo}
\end{figure}

We further investigate the effect of including HLOs on the QP correction to valence states. Figure \ref{fig:CuCl-g0w0-degw-hlo} shows the dependence of the QP correction to two particular valence states of CuCl on both $\nLO$ and $\dlLO$ in $G_0W_0$ calculation. The HLOs setups are the same for Cu and Cl for the sake of simplicity. By comparing the $G_0W_0$ self-energy corrections to the top valence band (dominantly Cu $3d$) and the 5th band below (mainly Cl $p$, denoted by VBM-5) at the $\Gamma$ point, we can see that the effect of including HLOs on the QP correction is associated with the characteristics of the corrected state, and is significantly larger for more localized $d$ states. 

\begin{table*}
    \centering
    \caption{Theoretical fundamental band gaps (eV) of cuprous and silver halides calculated by different theoretical approaches compared to experimental results. Data in the $\Delta_{\rm SO}$ column indicates the change in the fundamental gap when spin-orbit coupling (SOC) is considered, evaluated at the PBE level. The last two rows show the mean absolute error (MAE) of band gaps from different approaches compared to experimental data, without and with considering the effect of SOC.}
    \label{tab:band-gap}
    \begin{threeparttable}
        \begin{tabular}{lcccccccll}
            \hline \hline
            \multirow{2}*{Systems} & \multirow{2}*{PBE} & \multirow{2}*{HSE06} & \multicolumn{2}{c}{LAPW} &  \multicolumn{2}{c}{LAPW+HLOs} & {\multirow{2}*{$\Delta_{\rm SO}$}} & {\multirow{2}*{Previous $GW$}} & {\multirow{2}*{Expt.}} \\
            \cmidrule(lr){4-5}\cmidrule(lr){6-7}
            & & & $G_0W_0$ & $GW_0$ & $G_0W_0$ & $GW_0$ & & & \\
            \hline
            CuCl & 0.52 & 2.19 & 1.31 & 1.53 & 2.75 & 3.49 & $-$0.07 &  0.62\tnote{\it a}, 2.66\tnote{\it b}, 3.42\tnote{\it d} & 3.3990\tnote{\it g}, 3.2052\tnote{\it h}, 3.395\tnote{\it i} \\
            CuBr & 0.44 & 2.01 & 1.15 & 1.32 & 2.45 & 3.09 & $-$0.03 & 0.64\tnote{\it a}, 2.38\tnote{\it b}, 1.5\tnote{\it c}, 3.07\tnote{\it d}, 2.9\tnote{\it e} & 3.0726\tnote{\it j}, 3.077\tnote{\it i} \\
            CuI  & 1.12 & 2.50 & 1.78 & 1.88 & 2.82 & 3.29 & $-$0.16 & 1.79\tnote{\it a}, 2.70\tnote{\it f} & 3.115\tnote{\it i} \\
            AgCl & 0.87 & 2.18 & 1.83 & 2.04 & 2.62 & 2.99 & $-$0.04 & 2.16\tnote{\it a}, 2.97\tnote{\it b}, 3.29\tnote{\it d} & 3.2476\tnote{\it k} \\
            AgBr & 0.63 & 1.82 & 1.50 & 1.67 & 2.11 & 2.40 & $-$0.09 & 2.05\tnote{\it a}, 2.51\tnote{\it b}, 2.64\tnote{\it d} & 2.7125\tnote{\it l} \\
            AgI  & 1.30 & 2.35 & 2.14 & 2.30 & 2.63 & 2.90 & $-$0.23 & 2.77\tnote{\it a} & 2.91\tnote{\it m}\\
          \hline
            MAE     & 2.25 & 0.89 & 1.45 & 1.28 & 0.50 & 0.15 & & & \\
            MAE(SOC)& 2.36 & 0.99 & 1.55 & 1.38 & 0.61 & 0.18 & & & \\
         \hline \hline 
        \end{tabular}
        \begin{tablenotes}
            \item[{\it a}] From $G_0W_0$@PBE with Godby-Needs plasmon-pole model (PPM), reference \olcite{vanSetten2017}
            \item[{\it b}] From $G_0W_0$@LDA with Hybertsen-Louie PPM, reference \olcite{Gao2018}
            \item[{\it c}] From $G_0W_0$@LDA, reference \olcite{vanSchilfgaarde2006}
            \item[{\it d}] From $G_0W_0$@LDA+$U$, with Hybertsen-Louie PPM, reference \olcite{Gao2018}
            \item[{\it e}] From QS$GW$, reference \olcite{vanSchilfgaarde2006}
            \item[{\it f}] From QS$GW$, reference \olcite{Pishtshev2017}
            \item[{\it g}] From one-photon absorption spectra at 2 K, reference \olcite{Saito1995}
            \item[{\it h}] From two-photon absorption spectra (TPA) at 4.2 K, reference \olcite{Bivas1972} 
            \item[{\it i}] From reference \olcite{Goldmann1977}
            \item[{\it j}] From TPA at 1.6 K, reference \olcite{Mattausch1978}
            \item[{\it k}] From resonant Raman scattering (RRS) at 1.8 K, reference \olcite{Nakamura1983}
            \item[{\it l}] From RRS at 1.8 K, reference \olcite{Sliwczuk1984}
            \item[{\it m}] From optical spectra at 4 K, extracted from Fig. 1 of reference \olcite{Cardona1963}
        \end{tablenotes}
    \end{threeparttable}
\end{table*}

\subsection{Fundamental band gaps}\label{ss:fund-gap}
Table \ref{tab:band-gap} collects the calculated and experimental fundamental band gaps of all the cuprous and silver halides investigated. As expected, PBE underestimates the band gaps of all systems by more than 1.6 eV, with the largest discrepancy of 2.9 eV for CuCl. The generally more accurate HSE06 hybrid functional gives results in better agreement with experiment than PBE, but it is still not satisfactory with underestimation ranging from 0.6 to 1.2 eV. The results from PBE and HSE06 are consistent with the previous findings in the literature.\cite{Loftager2016, Pishtshev2017}

For $GW$ band gaps, we find that including HLOs in the LAPW basis leads to remarkable improvement for the band gap prediction for cuprous and silver halides. With the default LAPW basis, $G_0W_0$ gives an average quasi-particle correction to the band gap as 0.72 and 0.89 eV for CuX and AgX, respectively. Partial self-consistency of Green's function in $GW_0$ further opens the gap by 0.1 eV for CuI and 0.2 eV for CuCl, CuBr and all AgX. At this level of numerical accuracy, we can see that both $G_0W_0$ and $GW_0$ with PBE as the starting point performs unsatisfactorily for this class of materials. In particular, the $GW_0$ band gaps exhibit systematic underestimation errors in the range of 0.6 -- 1.7 eV, which are dramatically larger than typical errors observed in the same treatment of other semiconductors, and are even more severe for the well-known system ZnO. \cite{Jiang2016} When the LAPW+HLOs basis is used, we observe a significant increase in the $G_0W_0$ band gaps, averaged 1.26 and 0.63 eV for CuX and AgX, respectively. It is noted that the band gap increasing resulting from the inclusion of HLOs is more significant for the cuprous halides than silver halides, and increases in the order of iodide, bromide and chloride, which is consistent with previously found general trends that inclusion of HLOs have stronger effects on systems with more localized states and light elements.\cite{Jiang2016}

Obviously, by using LAPW+HLOs, PBE-based $GW_0$ can well predict fundamental band gaps of CuX and AgX with an mean absolute error (MAE) of about 0.15 eV, which is comparable to the errors of the same approach to typical $sp$ semiconductors. \cite{Jiang2016} The MAE of the $G_0W_0$ band gaps is 0.5 eV, which is still significantly smaller than those in previous reported results. Our investigation clearly indicates that physically CuX and AgX can still be regarded as ``simple'', i.e. weakly correlated, semiconductors, and that previous reported large errors in $GW$ calculation of these materials at the LDA/GGA-based $G_0W_0$ or $GW_0$ level can be mainly attributed to numerical inaccuracy.

When SOC is considered, the fundamental band gap is reduced due to the splitting of the top valence states $\Gamma_{15}$ for zincblende and ${\rm L}'_3$ for rocksalt systems. $\abs{\Delta_{\rm SO}}$ increases with larger atomic number of halogen, except for CuBr. This can be understood by the observation that splitting energy $\Gamma_8 - \Gamma_7$ is negative for CuCl but positive for CuBr and CuI.\cite{Cardona1963, Shindo1965} For all approaches investigated here, including SOC increases MAE. However, the magnitude of the increase is smaller for $GW_0$ with LAPW+HLOs than the other approaches, since the band gaps of CuX are slightly overestimated by $GW_0$ with LAPW+HLOs and the negative $\Delta_{\rm SO}$ reduces the errors.

To close this part, we make some remarks on the differences between our results and previously reported $GW$ results of CuX and AgX. In previous studies, LDA/GGA-based $G_0W_0$ were reported to underestimate the band gaps of CuX and AgX dramatically. In particular, van Setten \textit{et al.} performed a $G_0W_0$@PBE study with the Godby-Needs plasmon-pole model (PPM) and found that CuX are among the compounds that exhibit the largest errors in a high-throughput $GW$ study of a large set of insulating solids.\cite{vanSetten2017} They obtained fundamental band gaps of CuCl and CuBr of only 0.62 and 0.64 eV, respectively, which are about 0.5 eV smaller than those from $G_0W_0$ with default LAPW basis in the present study. Our $G_0W_0$ gap for CuBr with the standard LAPW is very close to that reported in Ref.\citenum{vanSchilfgaarde2006} that was also calculated in an all-electron $GW$ implementation. Meanwhile, it is worth noting that a recent work revealed that for molecular systems, the differences between results obtained from local orbital-based and plane-wave-based $G_0W_0$ implementations are greater for molecules containing Cu than other systems, which was attributed to the choice of pseudopotentials used in plane-wave-based implementation.\cite{Govoni2018-GW100} We thus suspect the dramatic errors in the band gaps of CuCl and CuBr by $G_0W_0$@PBE reported in Ref. \citenum{vanSetten2017} can be partly attributed to the inaccuracy of the pseudopotentials used in their study. For the band gaps of CuX, good agreement with experimental results has been obtained by using the self-consistent $GW$ (scGW) approaches. \cite{vanSchilfgaarde2006, Pishtshev2017, Azhikodan2017} However, as suggested by a series of careful studies, \cite{Shishkin2007, Cao2017, Grumet2018} different variants of scGW without considering vertex correction all tend to systematically overestimate the band gaps of typical semiconductors. The apparently good agreement between scGW results with experiment for CuX can be attributed to the error cancellation between the general tendency of scGW to overestimate the band gap and the numerical errors of $GW$ implementations based on the standard LAPW basis, as in Ref.\citenum{vanSchilfgaarde2006} or the use of pseudopotentials that tend to underestimate the band gap for such systems like CuX and ZnO.  

\begin{figure*}[!htp]
    \centering
    \includegraphics[width=0.3\linewidth]{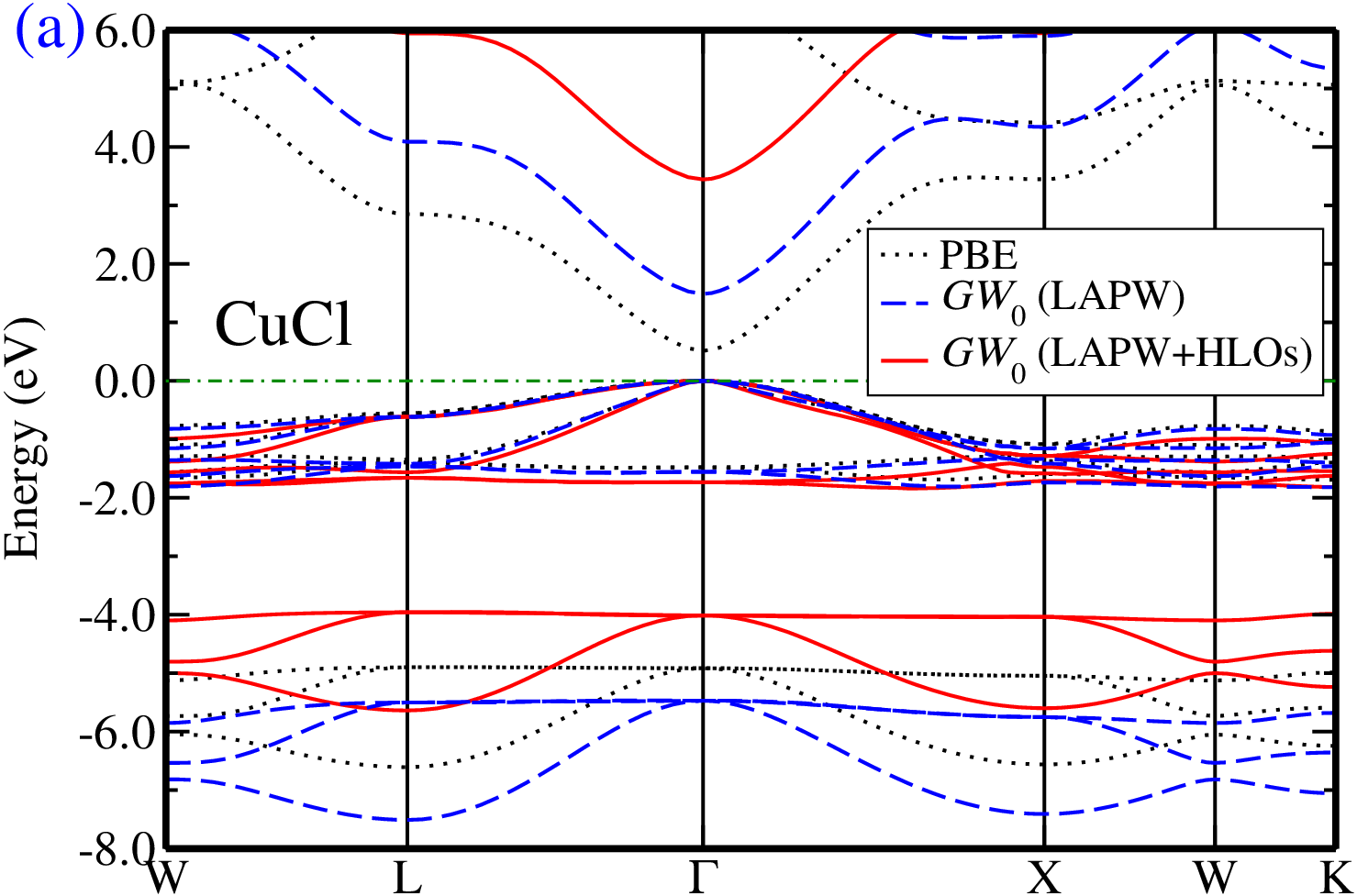}
    \includegraphics[width=0.3\linewidth]{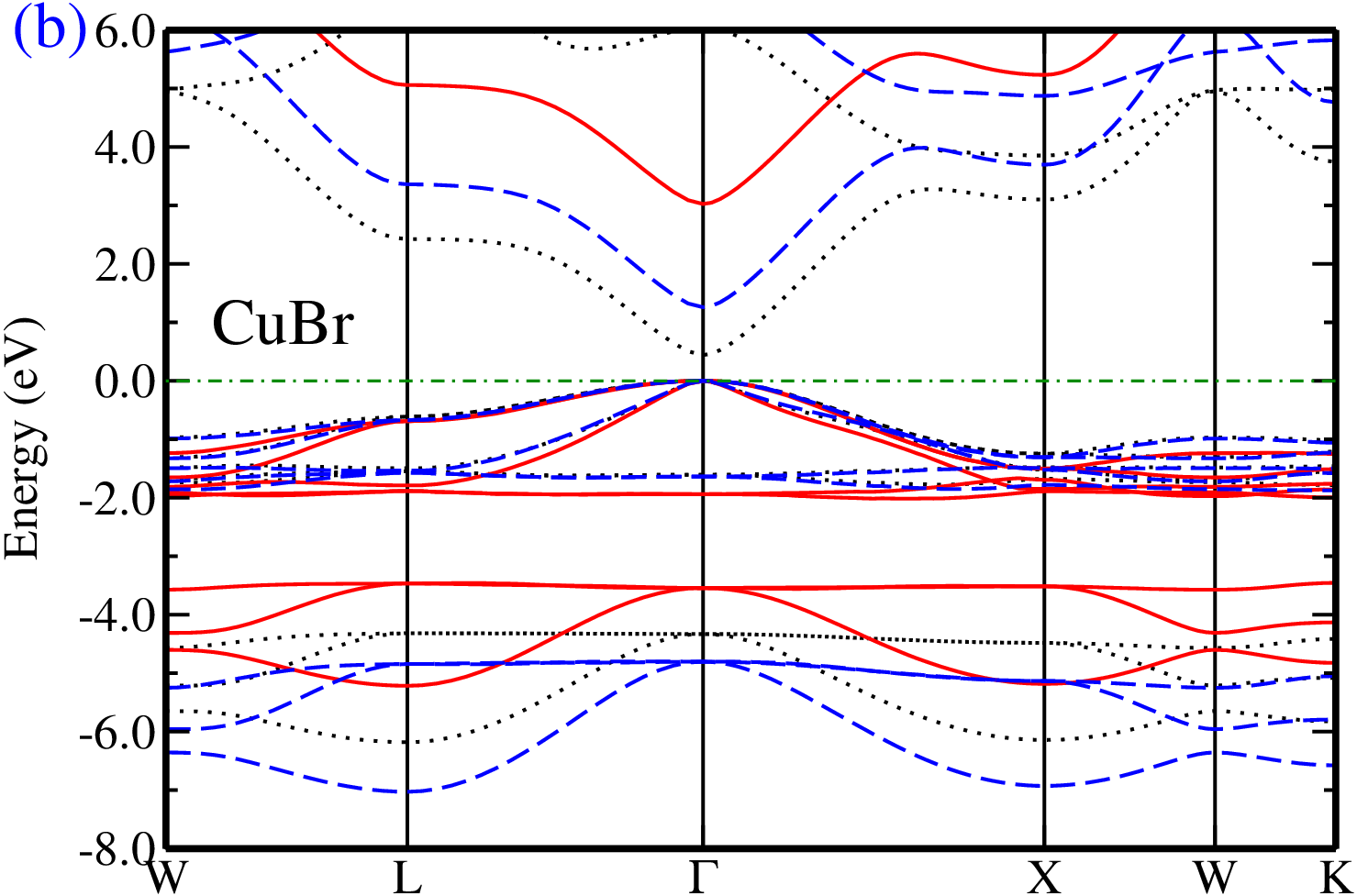}
    \includegraphics[width=0.3\linewidth]{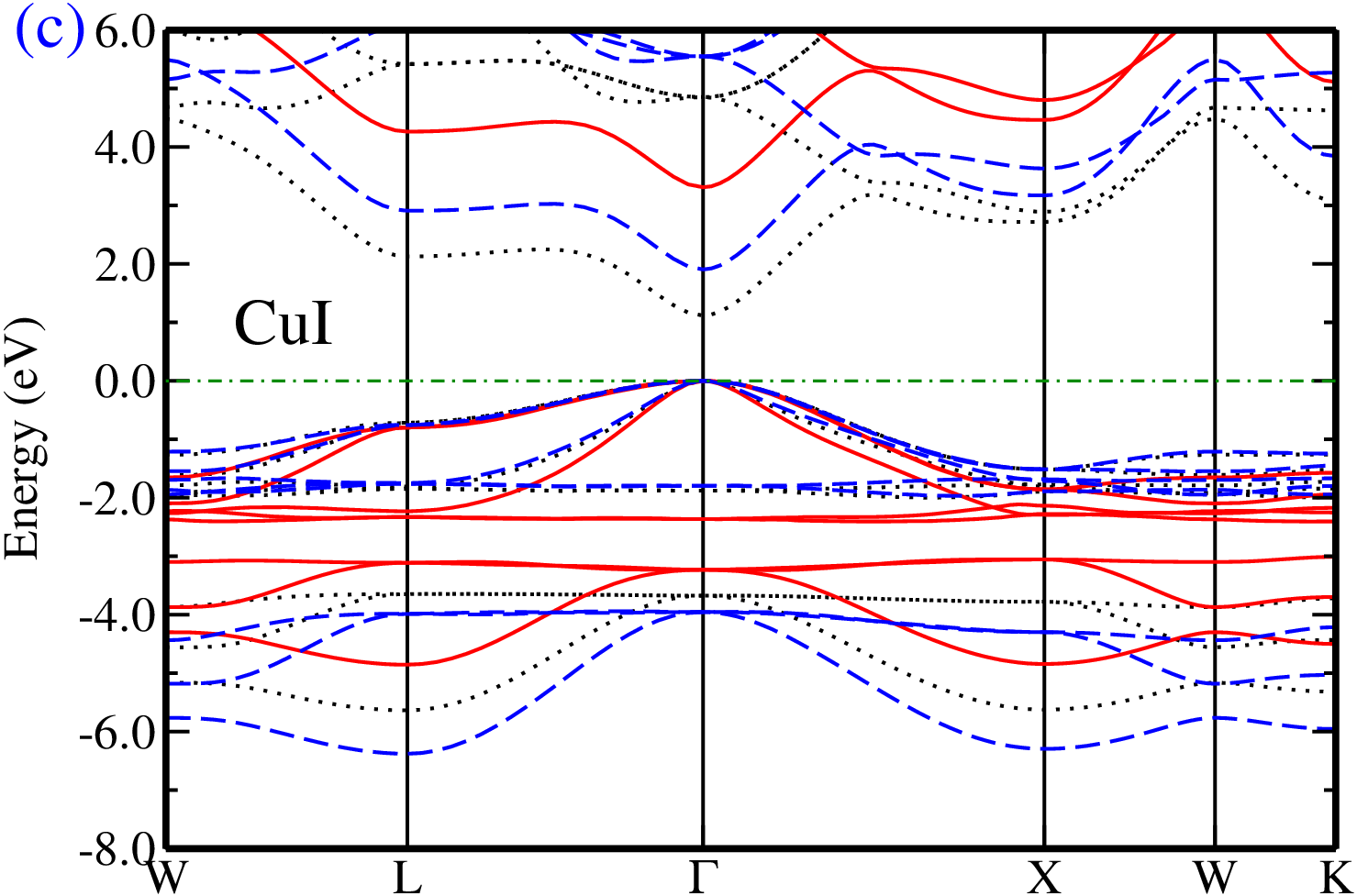}
    \\
    \includegraphics[width=0.3\linewidth]{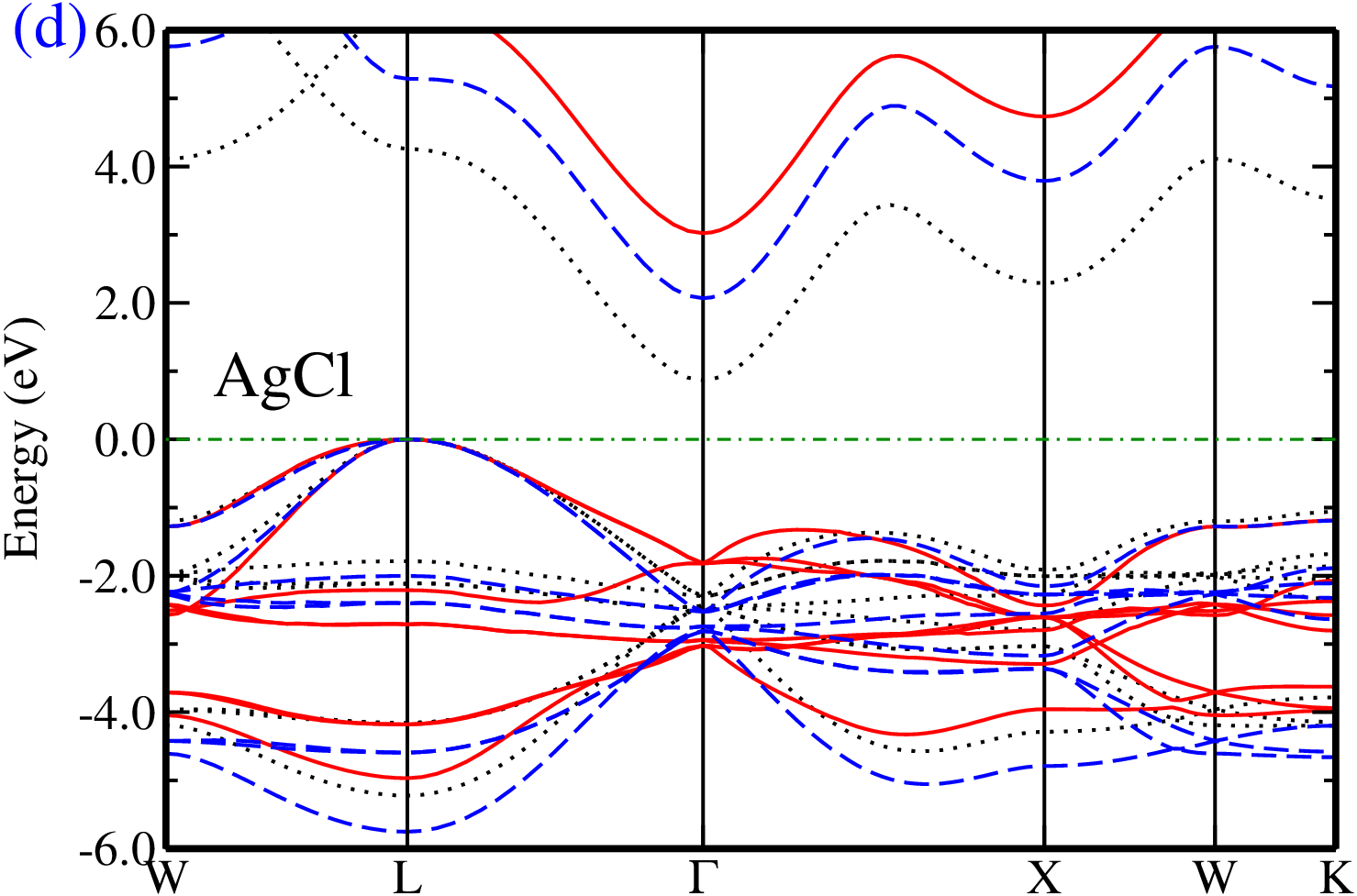}
    \includegraphics[width=0.3\linewidth]{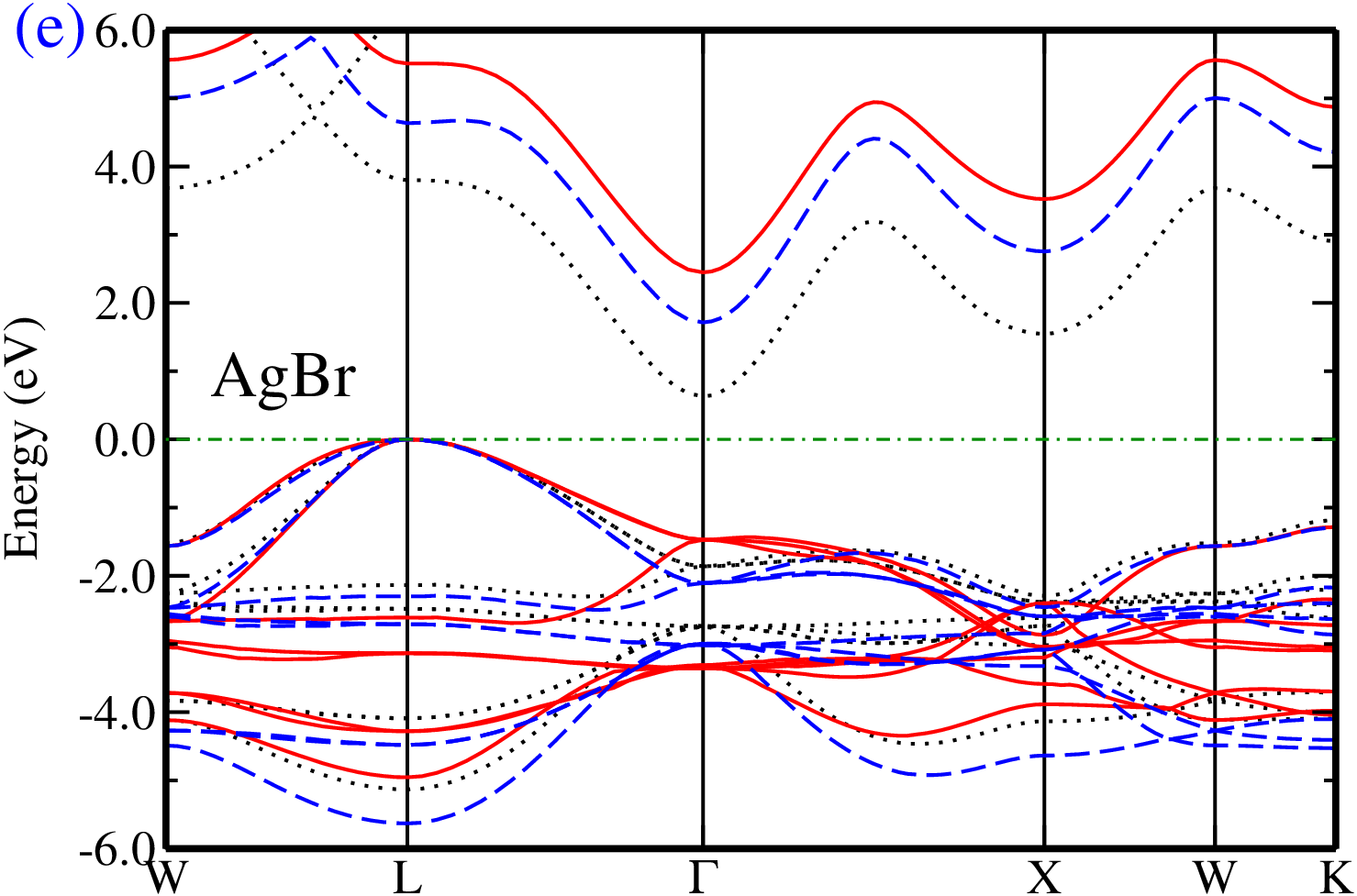}
    \includegraphics[width=0.3\linewidth]{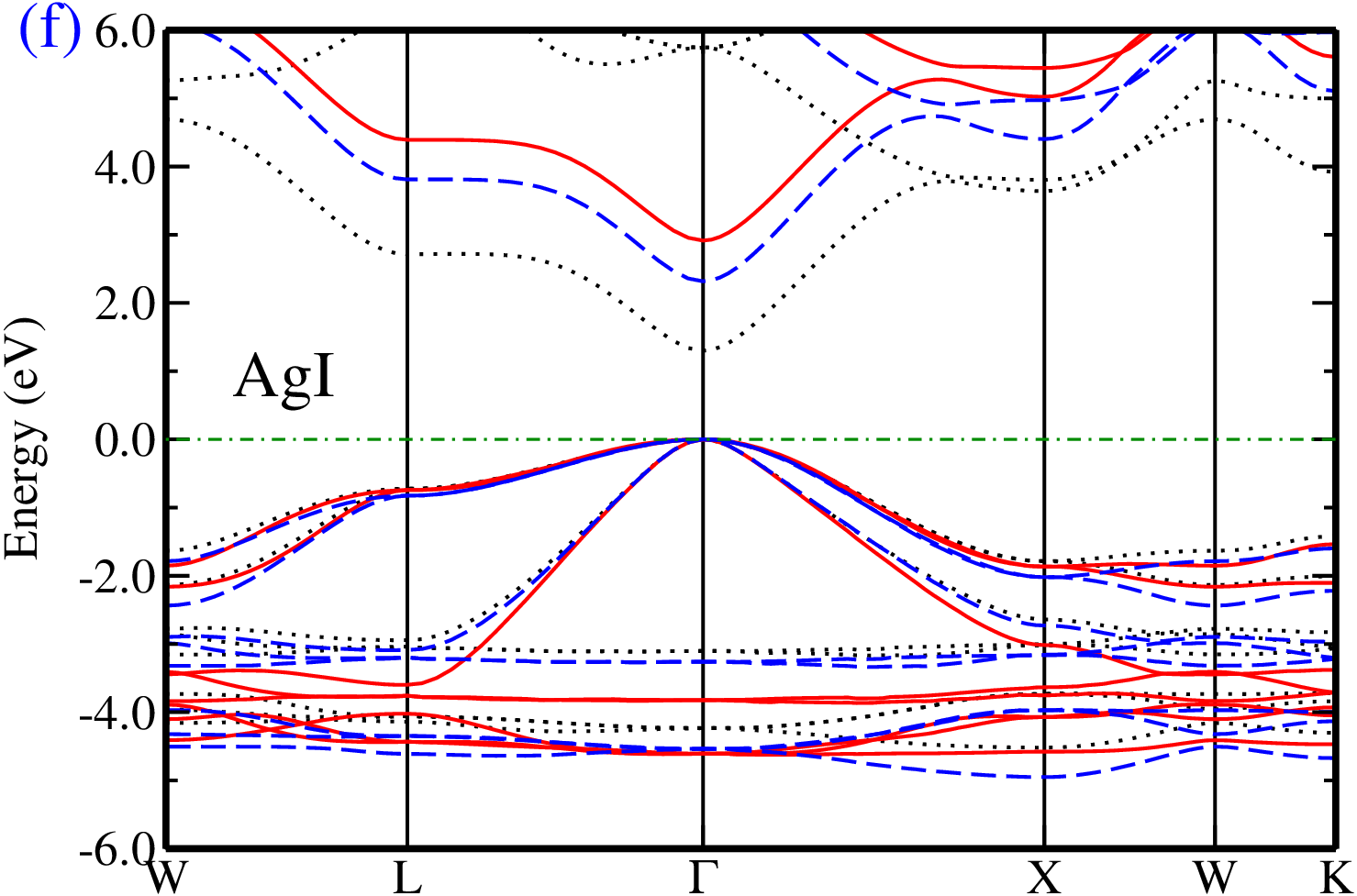} \\
    \caption{\ Electronic band structure near the Fermi level for cuprous and silver halides. The black dotted, blue dashed and red solid lines represent the energy bands calculated from PBE, $GW_0$(LAPW) and $GW_0$(LAPW+HLOs), respectively. The valence band maximum is aligned as the energy zero marked by the green dash-dotted line.}
    \label{fig:band-struct}
\end{figure*}

\subsection{Band structure and density of states}
We analyze in more details the effect of HLOs on the $GW$ calculation for cuprous and silver halides by scrutinizing the band structure diagrams of CuX and AgX as shown in Fig. \ref{fig:band-struct}. The energy zero is set to the valence band maximum for each case. We first discuss the features of PBE band structures of cuprous and silver halides. It is clearly seen that the systems with the zincblende structure, i.e. cuprous halides and AgI, have a direct minimal band gap at the $\Gamma$ point, while the systems with the rocksalt structure, i.e. AgCl and AgBr, have an indirect minimal band gap from L to $\Gamma$.  For cuprous halides, the three valence bands near the Fermi level are mixture of dominant Cu $3d$ $t_2$ ($d_{xy}$,$d_{yz}$,$d_{xz}$) and halide $np$ states ($n=3,4,5$ for X=Cl,Br,I, respectively), as suggested by the analysis of a quasi-molecular approach.\cite{Generalov2013} The relatively flat bands near $-2.0$ eV are almost exclusively formed by Cu $3d$ $e$ ($d_{x^2{\rm -}y^2}$, $d_{z^2}$) states and well separated from those lying between $-$8.0 eV and $-$3.0 eV, which are composed of mainly X-$np$ states. As the atomic number of halogen increases, the dispersion of the top valence bands increases and the separation between the Cu $3d$ and X $np$ bands decreases, as previously reported. \cite{Wei1988} The almost vanishing $d$-$p$ separation in AgI can be explained in a similar way, as Ag $4d$ and I $5p$ atomic orbitals are energetically close to each other. For AgCl and AgBr, X-$np$ and Ag-4$d$ states mix with each other in the valence regime, except for the $\Gamma$ point due to symmetry restriction. CBM of CuX and AgX is mainly composed of Cu-$4s$ and Ag-$5s$ states, respectively.

\begin{figure}[!ht]
    \centering
    \includegraphics[width=0.72\linewidth]{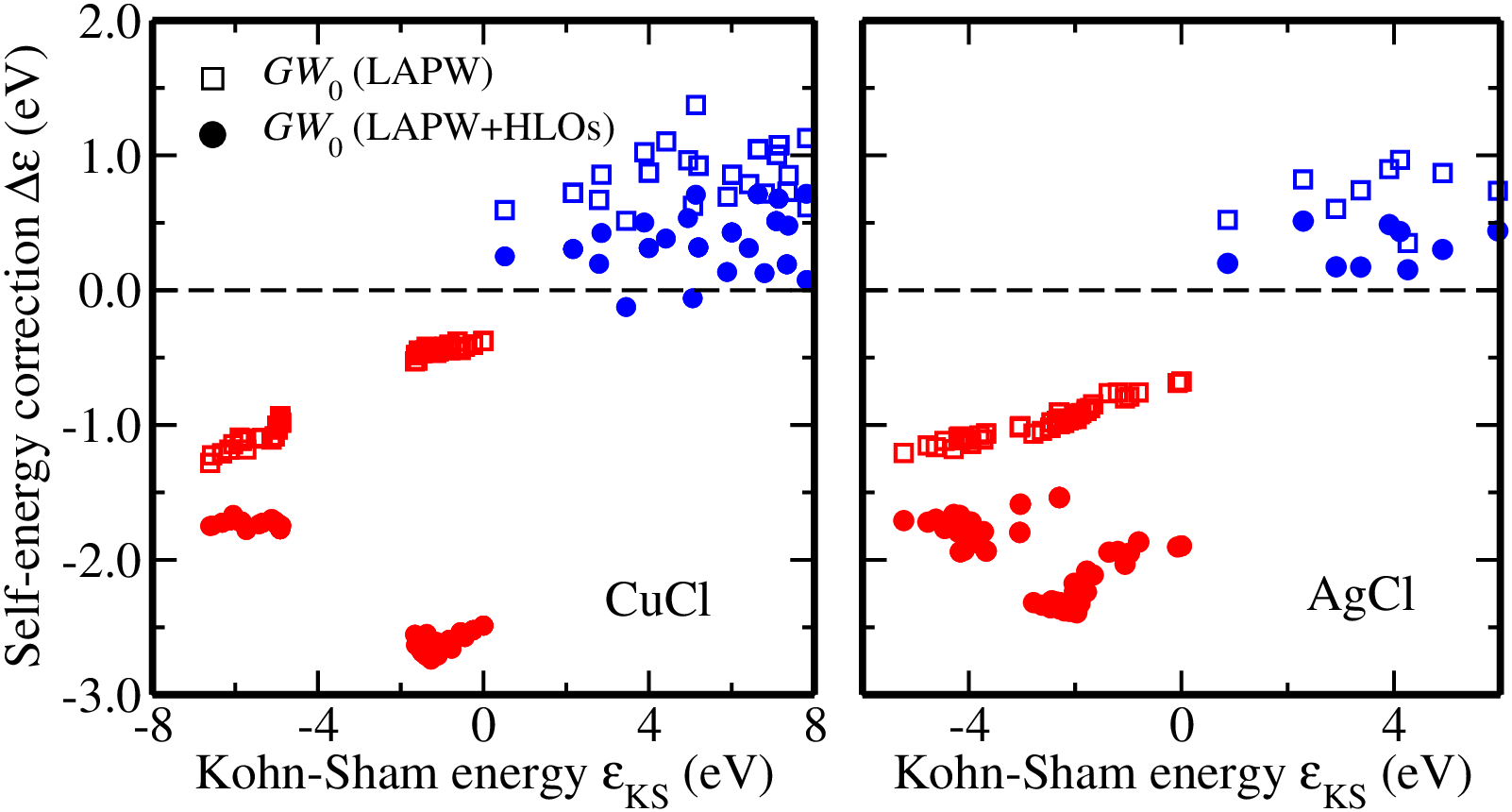}
    \caption{\ Self-energy or QP corrections to Kohn-Sham states calculated by $GW_0$ with standard LAPW basis (empty square) and LAPW+HLOs (solid circle). For a consistent comparison, the same setup of HLOs, $\nLO=5,\dlLO=4$, is used for the two systems. The color of red/blue indicates the valence/conduction states in CuCl (left panel) and AgCl (right panel). Zero correction is marked by the black dashed line.}\label{fig:degw-comparison}
\end{figure}

\begin{figure}[!ht]
    \centering
    \includegraphics[width=0.72\linewidth]{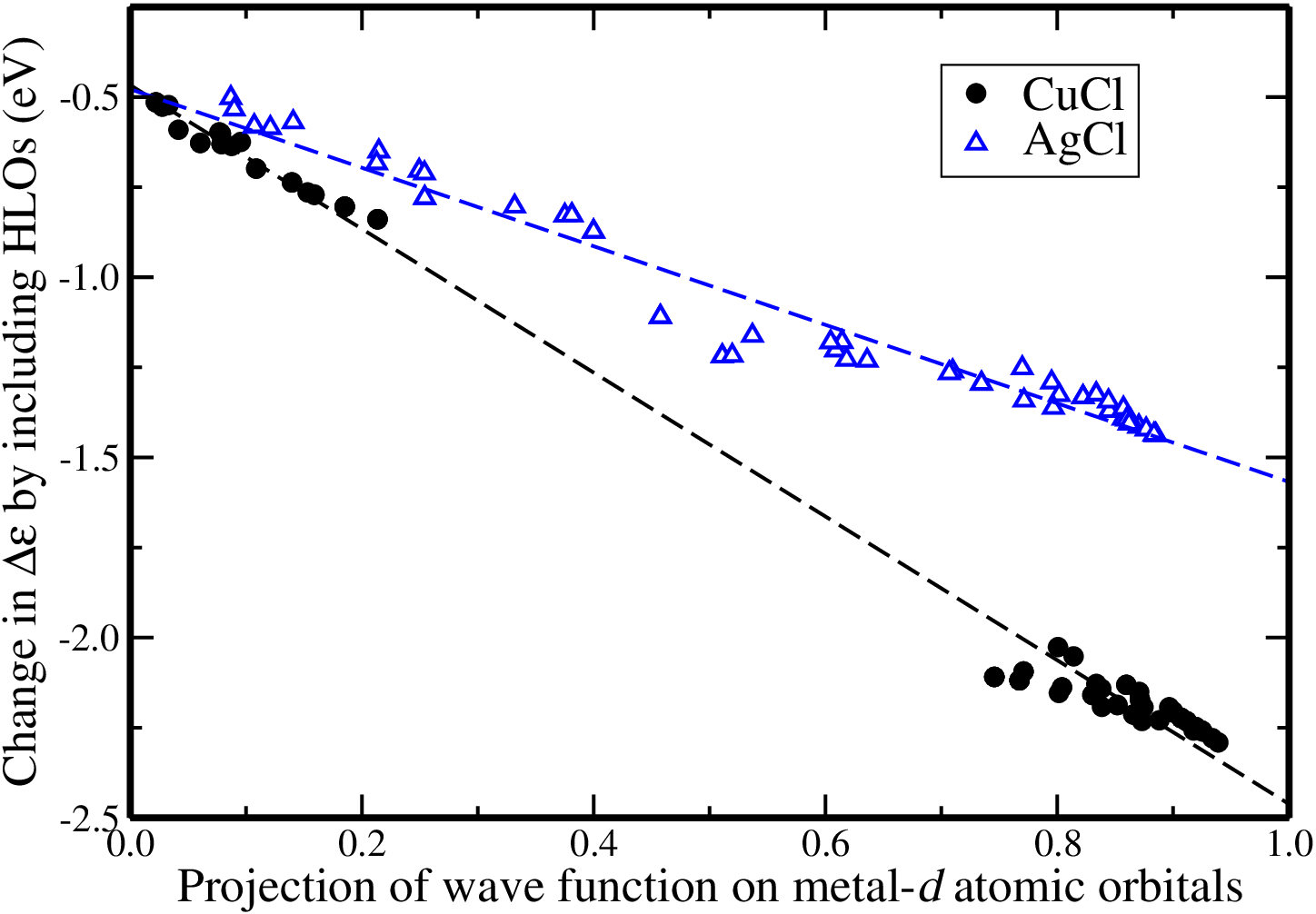}
    \caption{\ Dependence of the change in QP correction to valence band energies by $GW_0$@PBE method on the weight of metal-$d$ characters in the corresponding Kohn-Sham orbitals, $c_{n\veck}^{{\text{M-}}d}$, for CuCl (black circle) and AgCl (blue triangle) when HLOs ($\nLO=5,\dlLO=4$) is included in the basis set.}\label{fig:degw-pwave}
\end{figure}

\begin{figure*}[!ht]
    \centering
    \includegraphics[height=3.9cm]{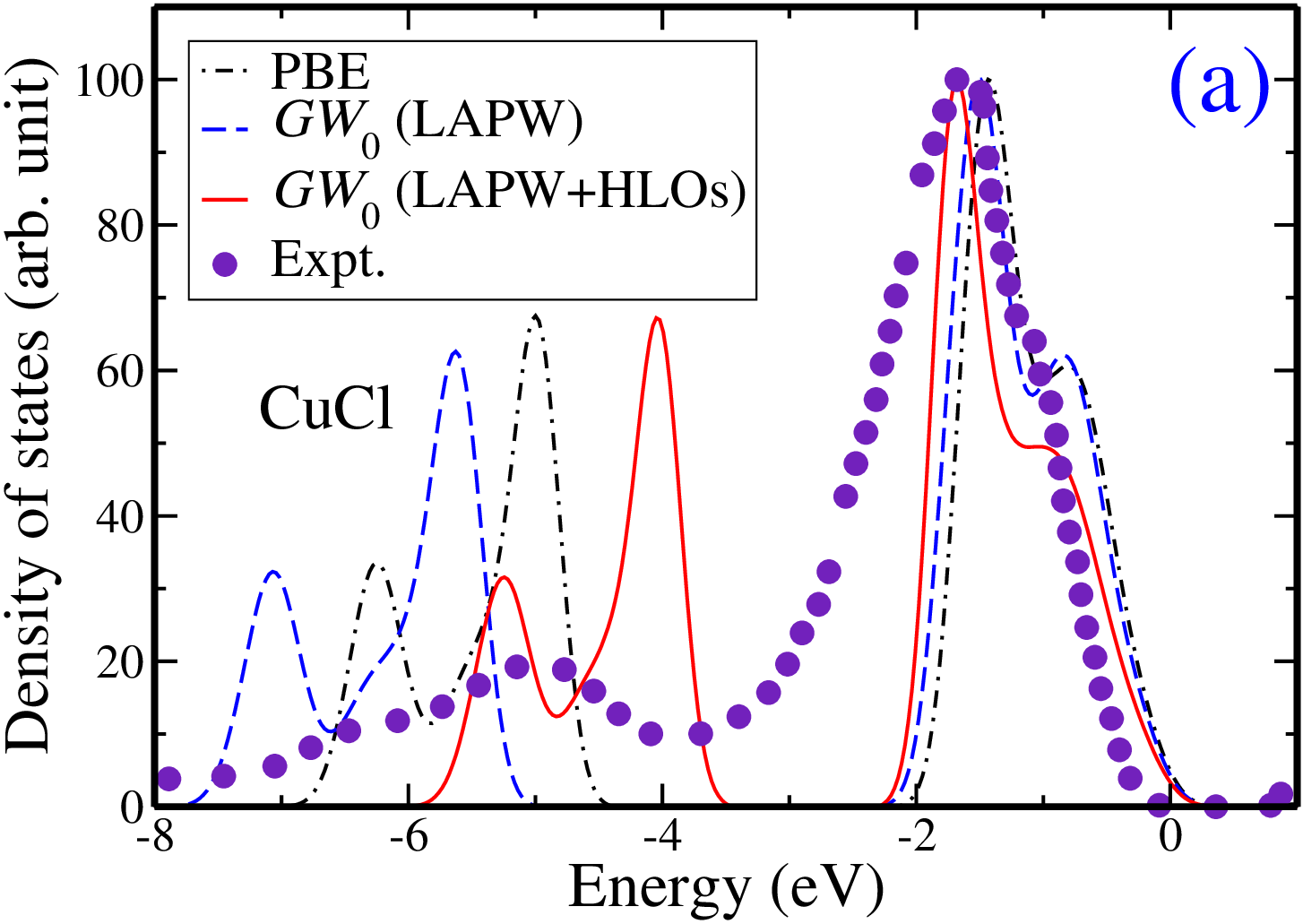}
    \includegraphics[height=3.9cm]{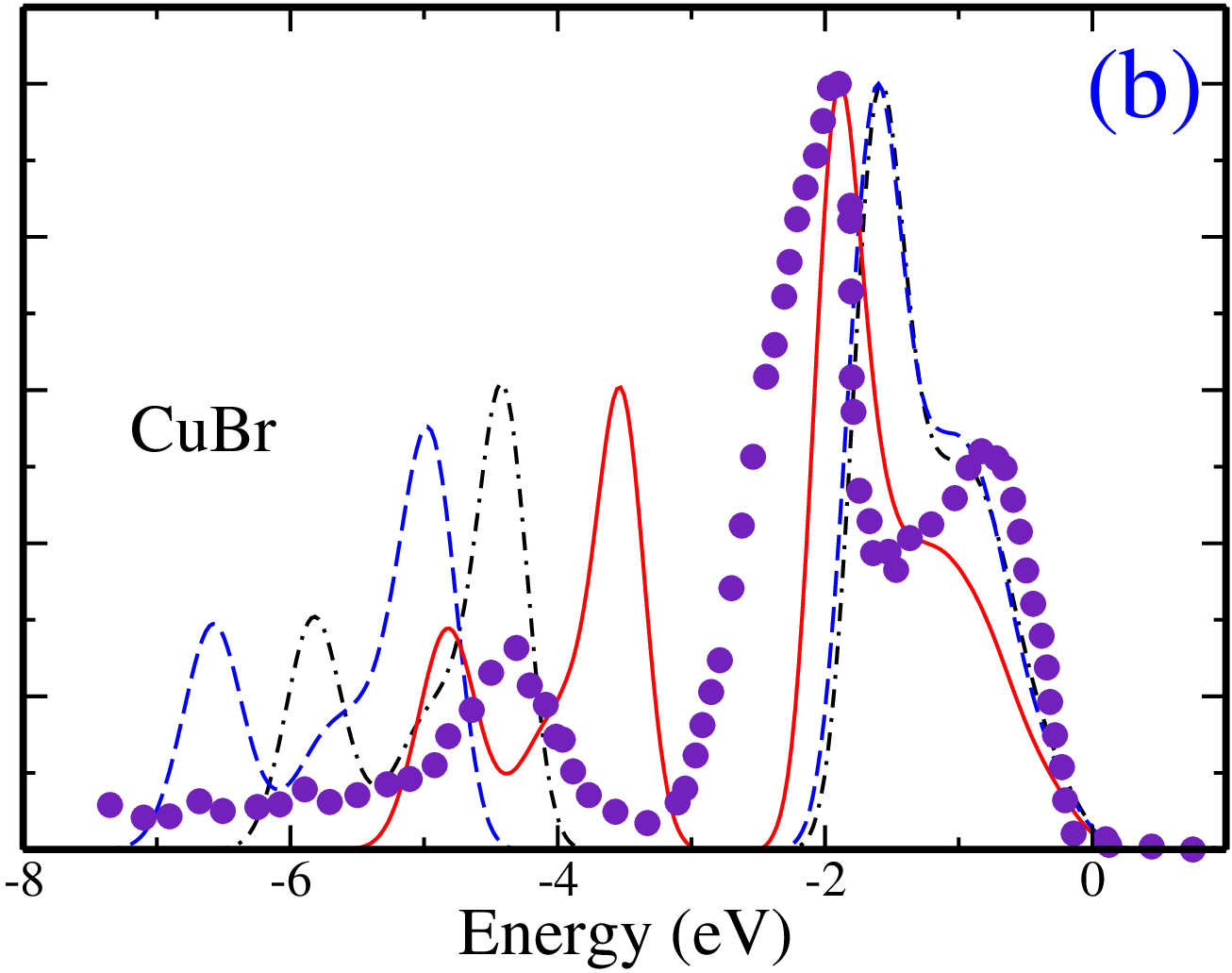}
    \includegraphics[height=3.9cm]{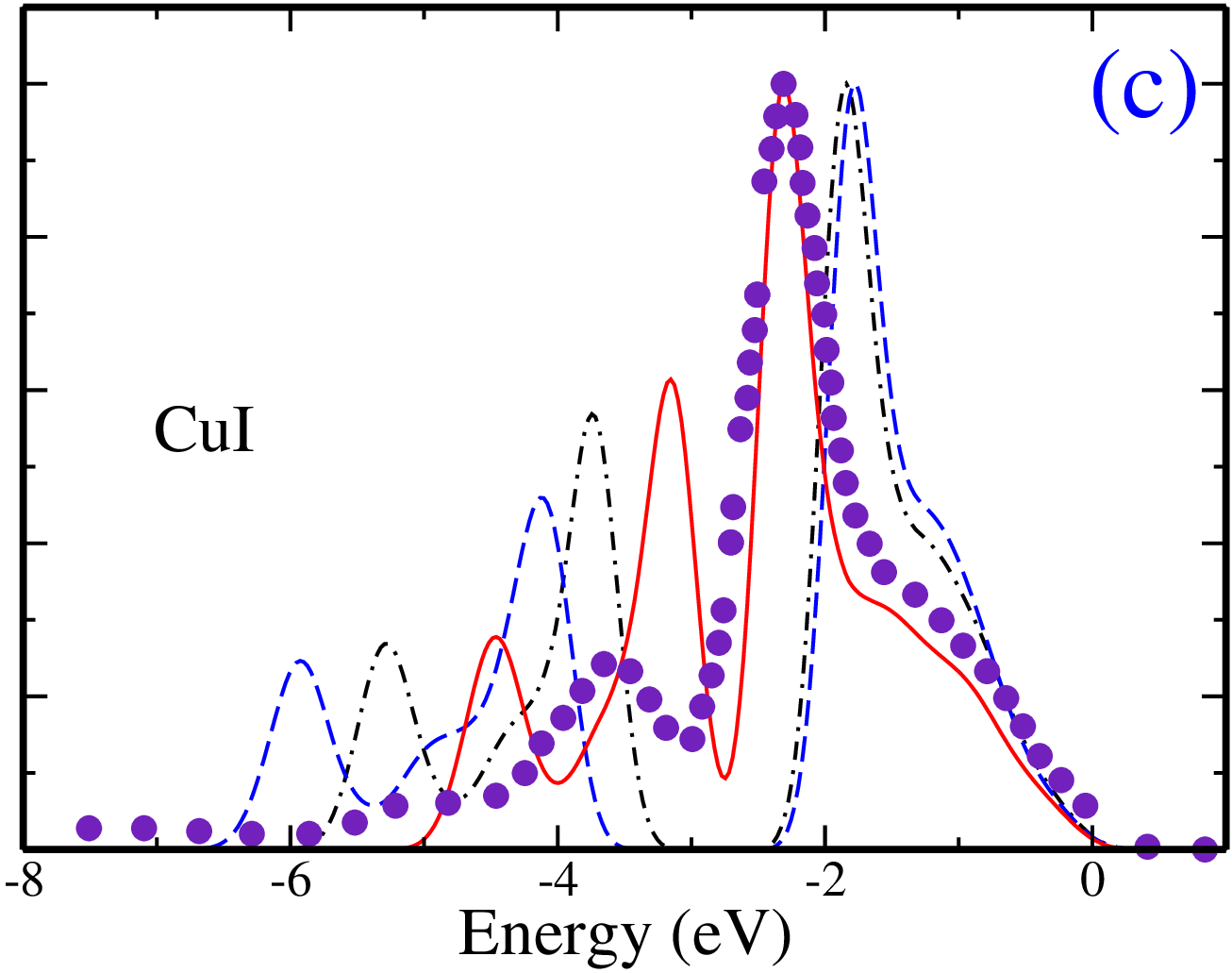} \\
    \includegraphics[height=3.9cm]{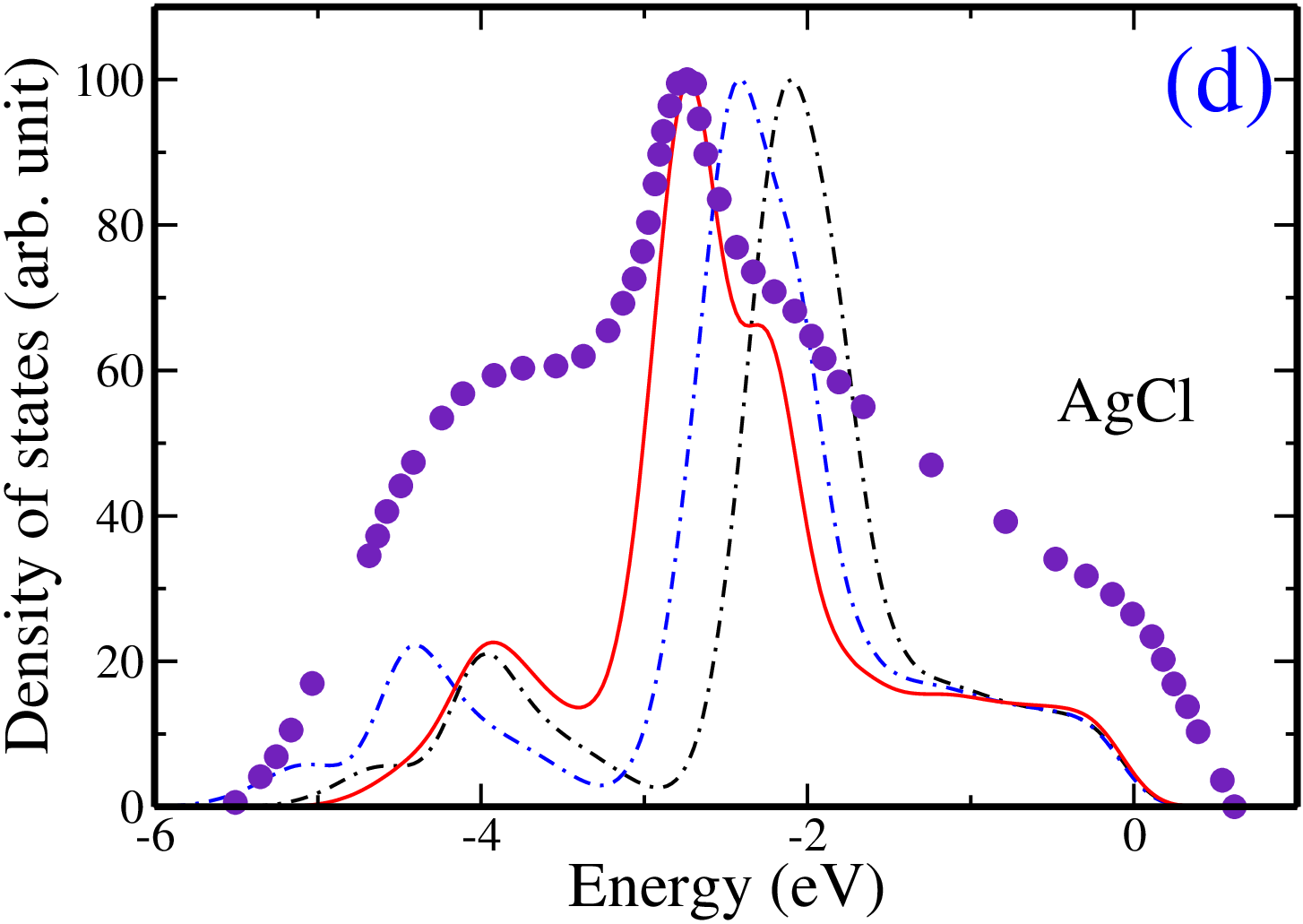}
    \includegraphics[height=3.9cm]{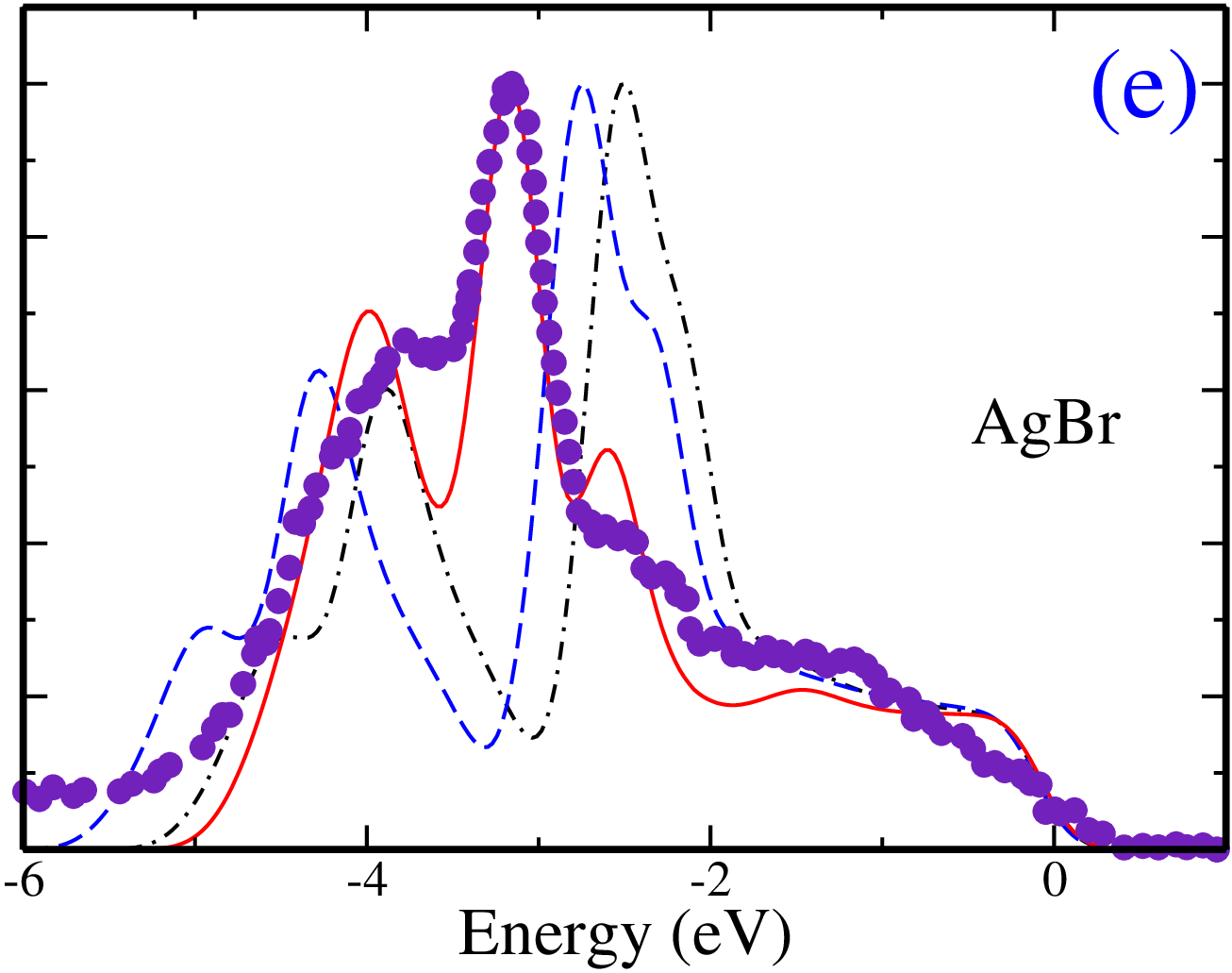}
    \includegraphics[height=3.9cm]{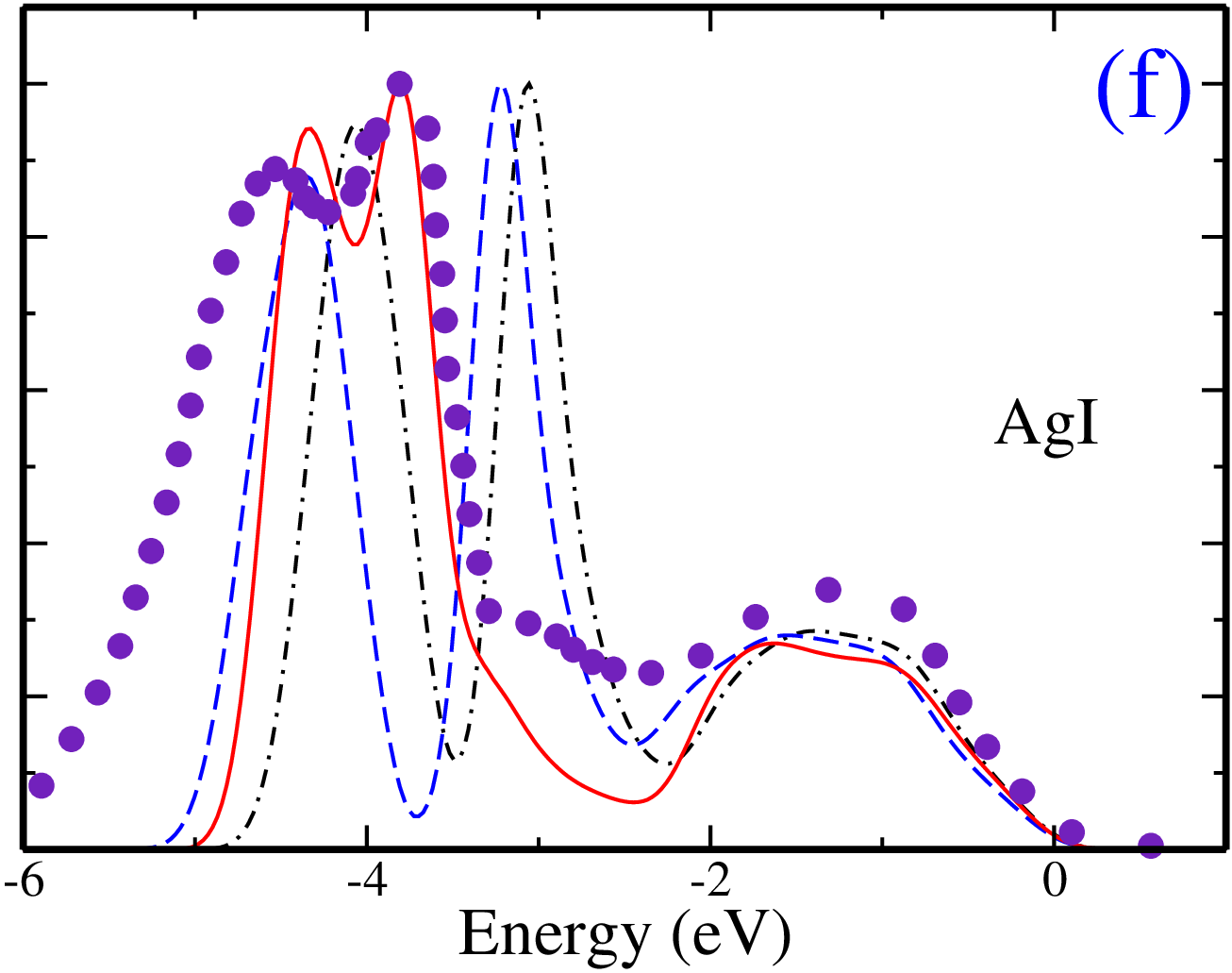} \\
    \caption{\ Calculated and experimental valence band density of states (DOS) for cuprous and silver halides. The black dotted, blue dashed and red solid lines represent the DOS calculated from PBE, $GW_0$ (LAPW) and $GW_0$ (LAPW+HLOs), respectively. The experimental spectral data (indicated by filled circles) for (a) and (c) is extracted from ref. \olcite{Generalov2013}, (b) from ref. \olcite{Matzdorf1993}, (d) from ref. \olcite{Tejeda1975}, (e) from ref. \olcite{Matzdorf1993a} and (f) from ref. \olcite{Goldmann1974}. Within each subgraph, the curves are normalized such that the strongest peaks have the same height. For convenient comparison, the original experimental data is rigidly shifted to make its strongest peak coincident with that of $GW_0$ (LAPW+HLOs), since uncertainties exist in experimentally determining the onset of electron emission and the Fermi level is defined differently from the theoretical one. The shifts for experimental data in subgraphs (a)-(f) are $-$0.450, $-$0.216, 0.037, 0.586, 0.409 and $-$0.768 eV, respectively.}
    \label{fig:dos}
\end{figure*}

Using PBE as the reference, we compare the band energies calculated by different methods. It is noted that PBE generally gives the right dispersion for valence states, while the band gaps are systematically underestimated. $GW_0$ with the default LAPW basis opens the band gap. Meanwhile, the energies of bonding $p$ bands in zincblende and $d$-$p$ band in rocksalt structures are pulled down with respect to the Fermi level. When comparing the band structures calculated from $GW$ with the default LAPW and LAPW+HLOs basis sets, we find that besides a greatly opened energy gap, the inclusion of HLOs also leads to a reduction in the separation between the $d$ and $p$ valence bands, which is clearly shown in the band structures of cuprous halides (Fig. \ref{fig:band-struct}a-c). This can be interpreted as a result of biased effects on self-energy correction to Kohn-Sham states of different characteristics by HLOs. For example, the self-energy corrections to Kohn-Sham band energies of CuCl and AgCl are presented in Fig. \ref{fig:degw-comparison}. When HLOs are included, the corrections to all states become more negative. However, the changes are more dramatic for valence states featuring metal $d$-characters than those with halogen $p$ and conduction states with metal $s$, leading to an enlarged band gap and narrowed $d$-$p$ separation. A more transparent picture can be obtained from Fig. \ref{fig:degw-pwave}, where the change in the QP correction to Kohn-Sham band energies when including HLOs is plotted against the weight of metal-$d$ characters in the corresponding Kohn-Sham orbital $\ket{\psi_{n\veck}}$, defined as 
\begin{equation}\label{eq:def-projection}
  c_{n\veck}^{\text{M-}d} = \sum_{m=-2}^2{\abs{\braket{\phi^{{\rm M}}_{l=2, m}}{\psi_{n\veck}}}^2}, 
\end{equation}
where $\ket{\phi^{{\rm M}}_{l=2, m}}$ is the predefined atomic-like basis featuring spherical harmonic function $Y^m_2$ within the muffin-tin of the M atom (M=Cu for CuX and Ag for AgX). More dramatic change in QP correction is observed for the valence state with larger $c_{n\veck}^{\text{M-}d}$. Furthermore, a linear regression of the change in QP correction to valence state by including HLOs with respect to $c_{n\veck}^{\text{M-}d}$ shows a similar intercept for CuCl and AgCl, but gives a slope for CuCl ($-$2.0 eV) almost two times larger than that for AgCl ($-$1.1 eV), indicating stronger effects of including HLOs on Cu-$3d$ than Ag-$4d$. The slopes for bromide and iodide are different from that of the corresponding chloride by less than 0.1 eV.

Finally we compare the density of states in the valence regime calculated by using different methods with that obtained from the photo-electronic spectroscopy experiments. As shown in Fig. \ref{fig:dos}, while significantly underestimating the band gap, PBE in general predicts the peak positions in valence spectral data in reasonable agreement with experiment. $GW_0$@PBE with the default LAPW basis overestimates the $d$-$p$ separation systematically. For example, the peaks of Cu 3$d$ and Br 4$p$ bands in CuBr are separated by 3.4 eV, almost 1 eV larger than the experimental value of about 2.4 eV. Such discrepancy is resolved by $GW_0$ with LAPW+HLOs, which gives accurate peak separation for silver halides, but slightly underestimates the splitting for cuprous halides compared to experiment.  

\section{Conclusions}\label{sec:conclusion}

Previous LDA/GGA-based $G_0W_0$ calculations have confronted difficulties in accurately predicting the quasi-particle band structure of CuX and AgX (X=Cl, Br, I).
In this paper, we have performed the $G_0W_0$ and $GW_0$ calculations from PBE input for these materials based on the all-electron implementation with LAPW basis extended by high-energy local orbitals (HLOs). It is demonstrated that not only the band gaps, but also the separations between $d$ and $p$ bands in the valence regime are predicted in close agreement with the experiments.
Both facts stem from a biased correction to self-energy of states with different atomic characteristics by including HLOs in the basis set. Within the same system, larger corrections are generally observed in energy states with greater metal $d$ components, and hence it is crucial to include HLOs in order to accurately evaluate the self-energy correction to the localized $d$ states.
Moreover, we show that self-energy corrections to Cu $3d$ states are more sensitive to the inclusion of HLOs than those to Ag $4d$ by linearly regressing the change in self-energy correction by HLOs with respect to the projection of wave function on $d$ atomic orbitals for the valence states. We have also performed a detailed convergence test of quasi-particle band gap with respect to the two controlling parameters of HLOs, namely $\nLO$ and $\dlLO$.
Systematically added HLOs centered on Cu and Ag atoms brings much more correction on the quasi-particle band gap than those on halogen atoms, which is exploited here to achieve a reasonable convergence level of band gaps without making the basis overwhelmingly large. Combining the current study on CuX and AgX and the previous one on ZnO,\cite{Jiang2016} we emphasize the highly system-dependent feature of the effect on the quasi-particle band structure of HLOs that vary rapidly near the nuclei, and its significance for theoretically describing the electronic and optical properties of materials containing $d^{10}$ transition metals.

\begin{acknowledgments}\label{sec:ack}
This work is partly supported by the National Natural Science Foundation of China (21673005, 21621061). The authors also acknowledge the support by High-performance Computing Platform of Peking University for the computational resources.
\end{acknowledgments}

\bibliography{CuAgX-GW}

\end{document}


\maketitle

This \emph{Supplemental Material} presents the results of convergence tests on the fundamental band gap $E_{\rm g}$ and quasi-particle (QP) correction $\Delta\varepsilon$ calculated within the $G_0W_0$@PBE framework for cuprous and silver halides, with respect to the parameters $\nLO$ and $\dlLO$ of the setup of high-energy local orbitals (HLOs). All tests are performed with $2\times2\times2$ $\Gamma$-centered $\bf k$-point mesh.

Note that the results of CuCl are presented in the main article for a detailed discussion and are not repeated here. 

\newpage

\beginsupplement

\begin{figure}[p]
    \centering	
    \includegraphics[width=0.9\linewidth]{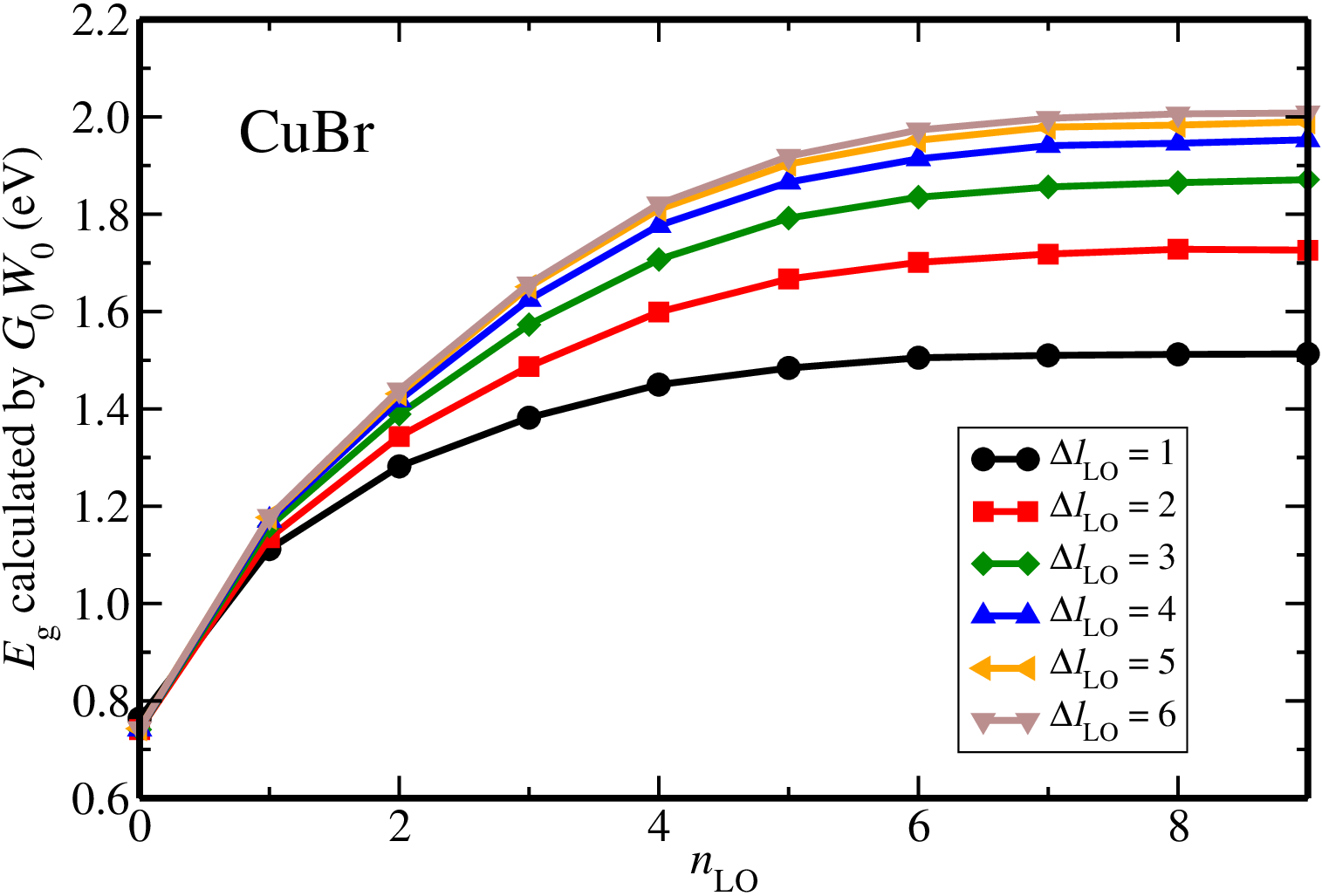}
    \caption{\ Fundamental band gap of CuBr calculated by $G_0W_0$@PBE with different HLOs setups, characterized by $\nLO$ and $\dlLO$.}
    \label{figsi:CuBr-g0w0-gap-vs-hlo}
\end{figure}

\begin{figure}[p]
    \centering	
    \includegraphics[width=0.9\linewidth]{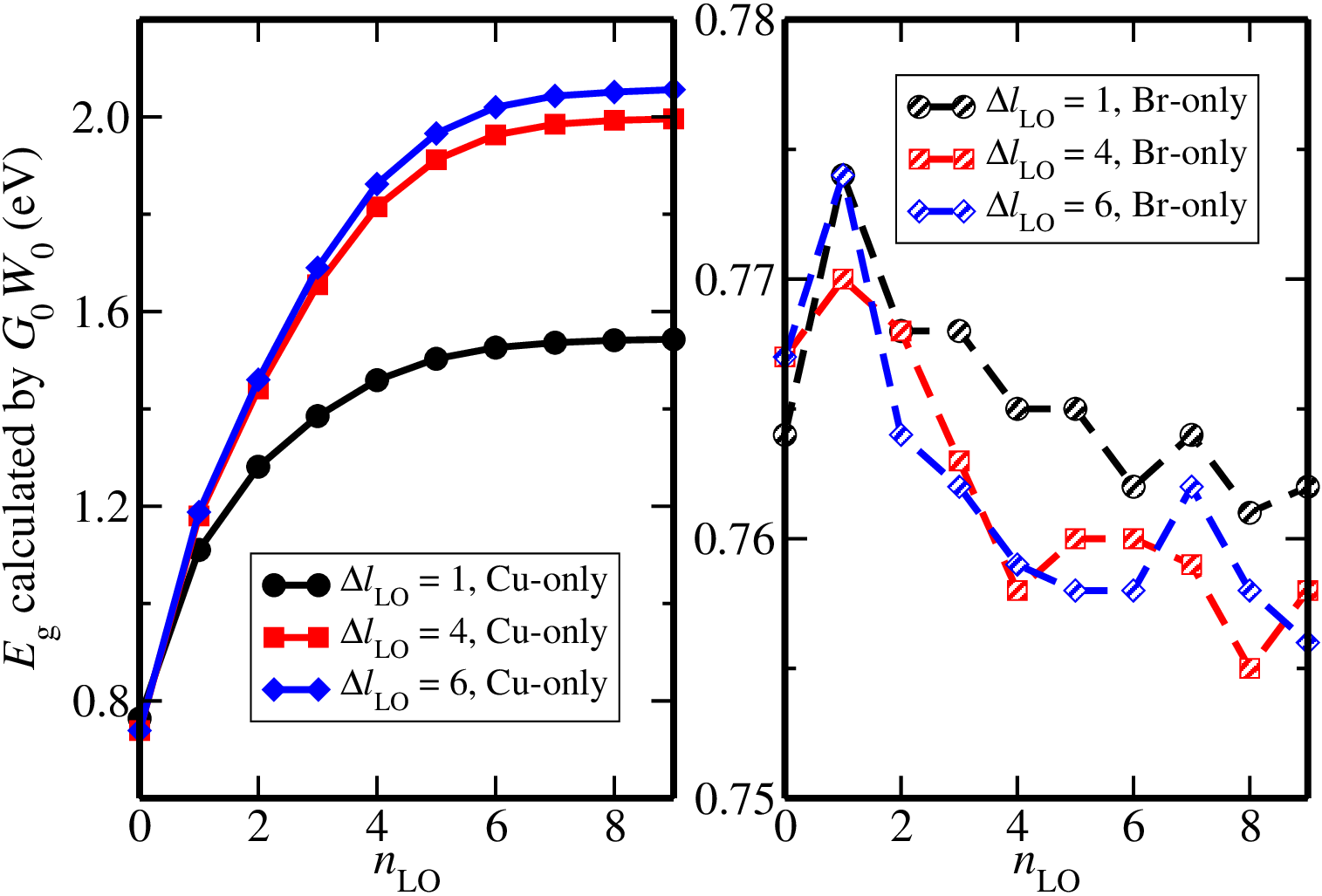}
    \caption{\ Fundamental band gap of CuBr calculated by $G_0W_0$@PBE with HLOs added on either Cu or Br atoms.}
    \label{figsi:elem-dependence-CuBr-g0w0-gap}  
\end{figure}

\begin{figure}[p]
    \centering
    \includegraphics[width=0.9\textwidth]{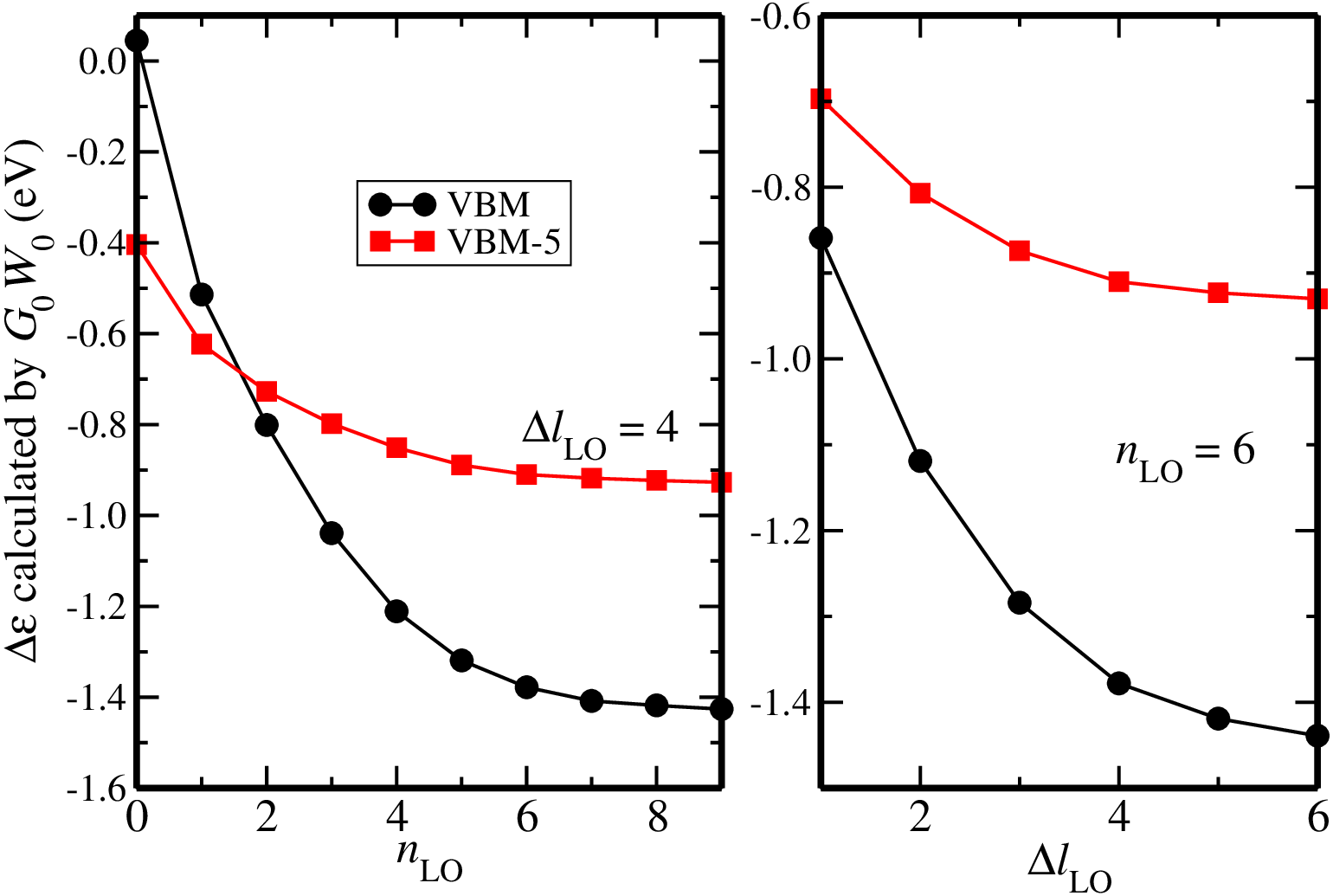}
    \caption{\ Dependence of $G_0W_0$ QP correction to the valence states at the $\Gamma$ point on $\nLO$ and $\dlLO$ of HLOs setup in CuBr. ``VBM'' denotes the top valence band and ``VBM-5'' the fifth band below VBM. $\nLO$ and $\dlLO$ are fixed as 6 and 4, respectively, when the other parameter is varying.}
    \label{figsi:CuBr-g0w0-degw-hlo}
\end{figure}

\newpage

\begin{figure}[p]
    \centering	
    \includegraphics[width=0.9\linewidth]{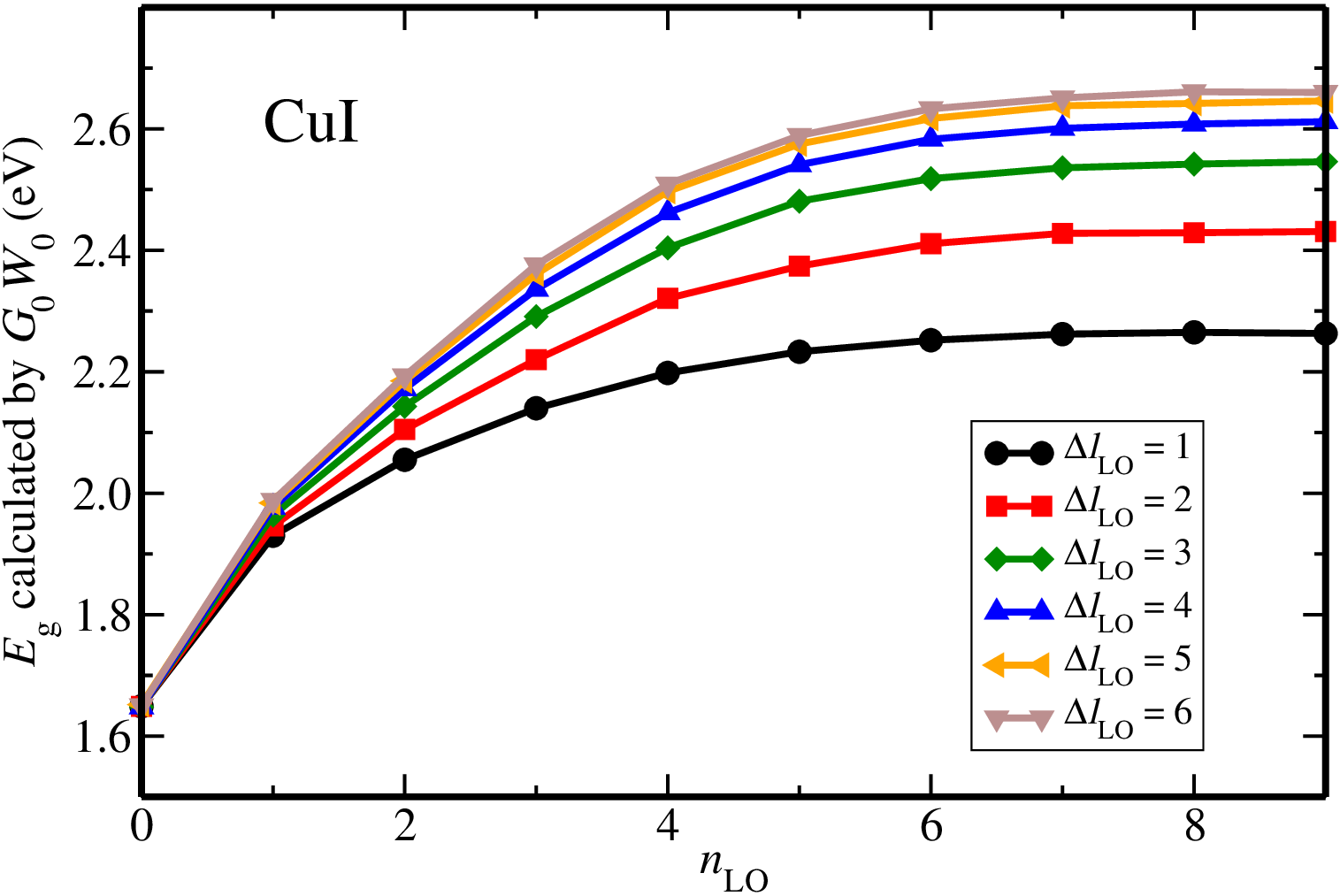}
    \caption{\ Similar to \ref{figsi:CuBr-g0w0-gap-vs-hlo}, but for CuI.}
    \label{figsi:CuI-g0w0-gap-vs-hlo}
\end{figure}

\begin{figure}[!ht]
    \centering	
    \includegraphics[width=0.9\linewidth]{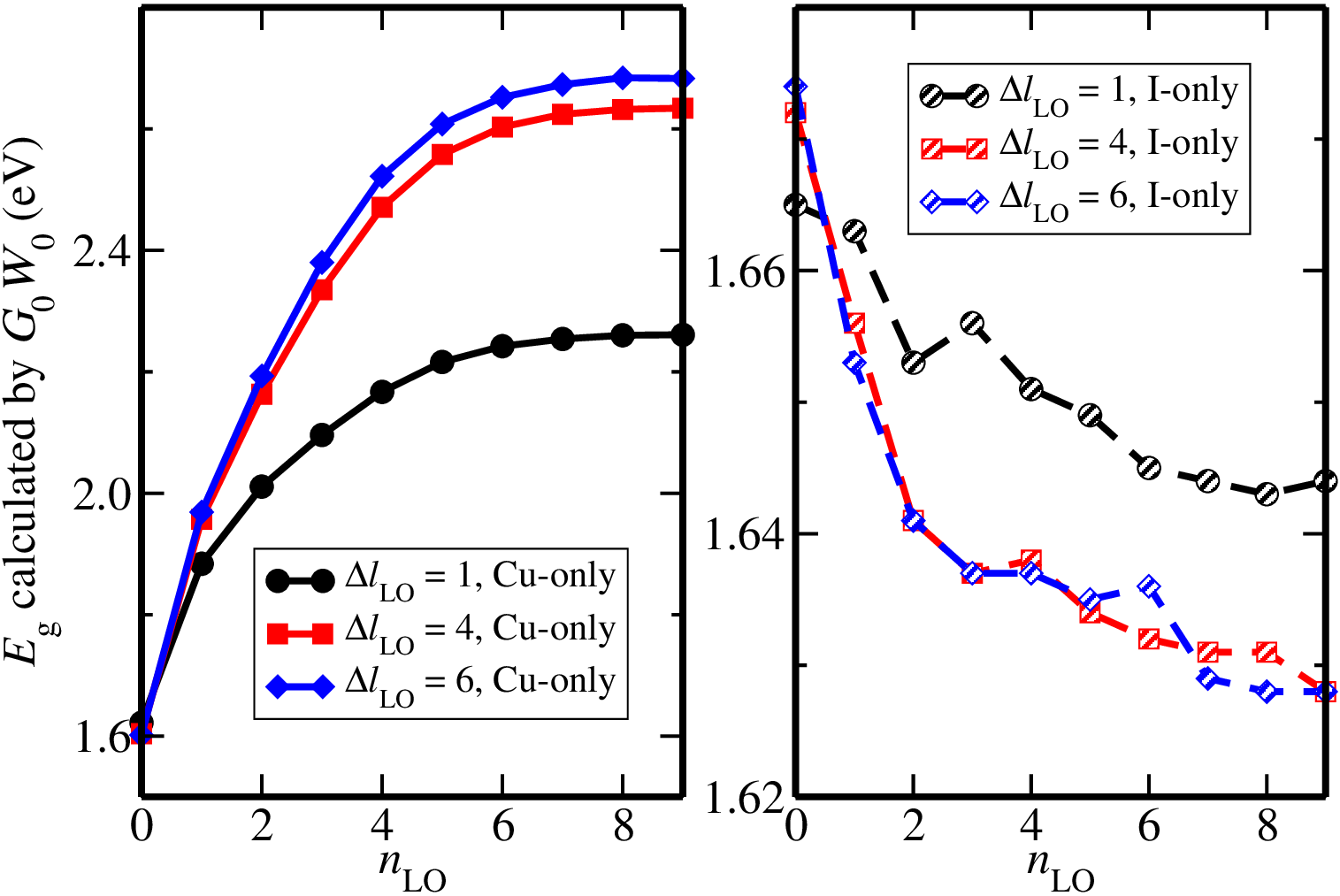}
    \caption{\ Fundamental band gap of CuI calculated by $G_0W_0$@PBE with HLOs added on either Cu or I atoms.}
    \label{figsi:elem-dependence-CuI-g0w0-gap}  
\end{figure}

\begin{figure}[!ht]
    \centering
    \includegraphics[width=0.9\textwidth]{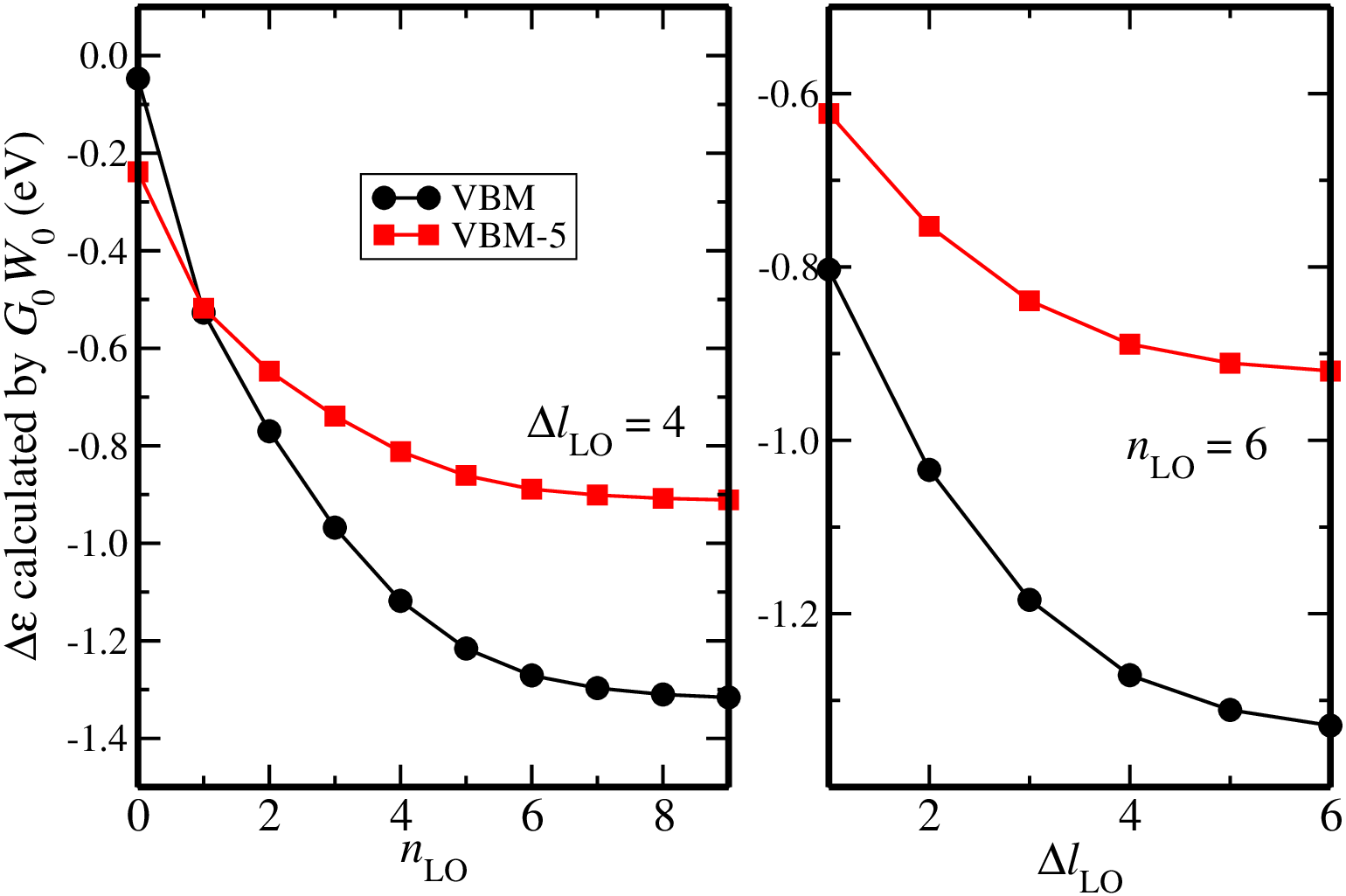}
    \caption{\ Similar to \ref{figsi:CuBr-g0w0-degw-hlo}, but for CuI.}
    \label{figsi:CuI-g0w0-degw-hlo}
\end{figure}

\newpage

\begin{figure}[!ht]
    \centering	
    \includegraphics[width=0.9\linewidth]{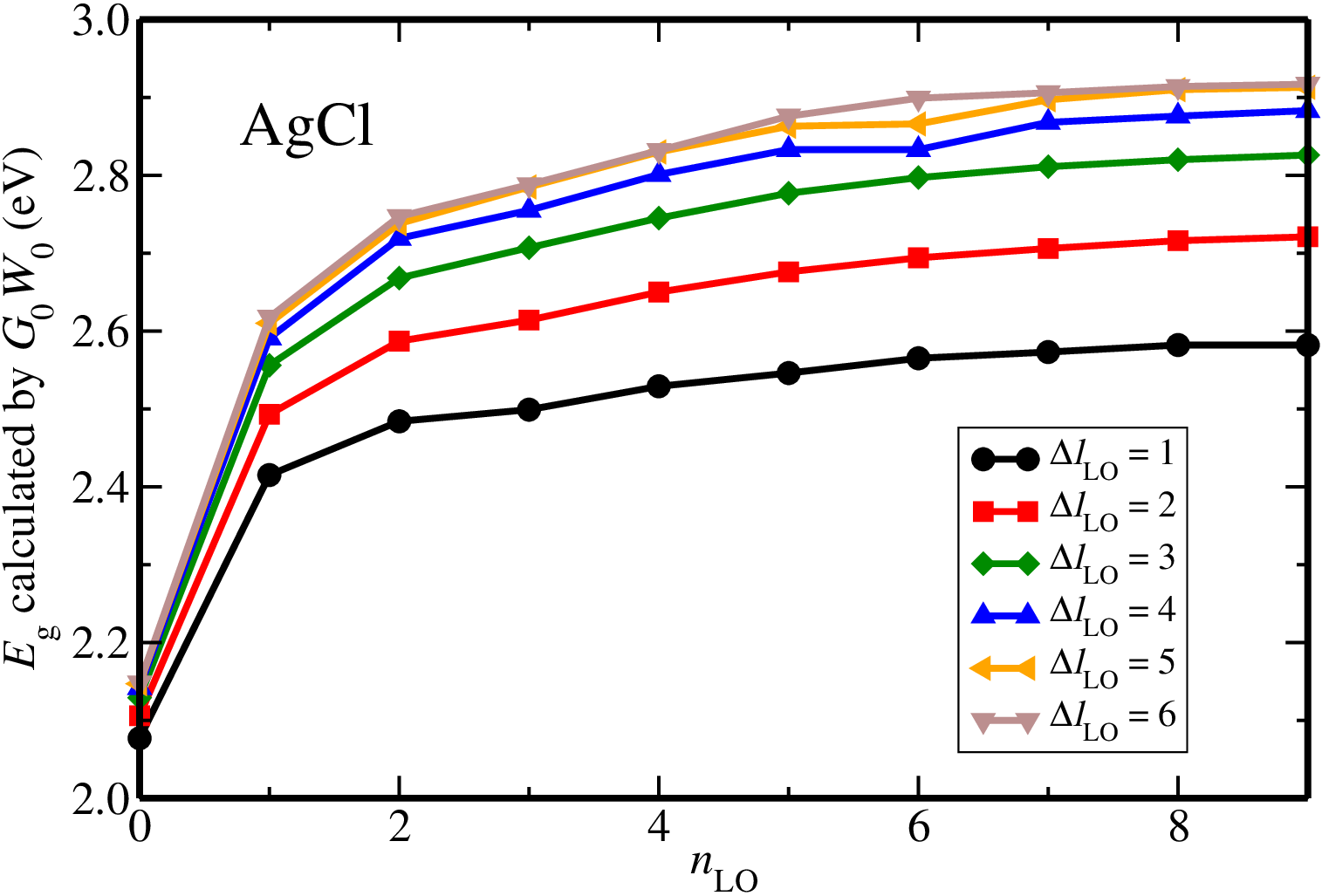}
    \caption{\ Similar to \ref{figsi:CuBr-g0w0-gap-vs-hlo}, but for AgCl}
    \label{figsi:AgCl-g0w0-gap-vs-hlo}
\end{figure}

\begin{figure}[!ht]
    \centering	
    \includegraphics[width=0.9\linewidth]{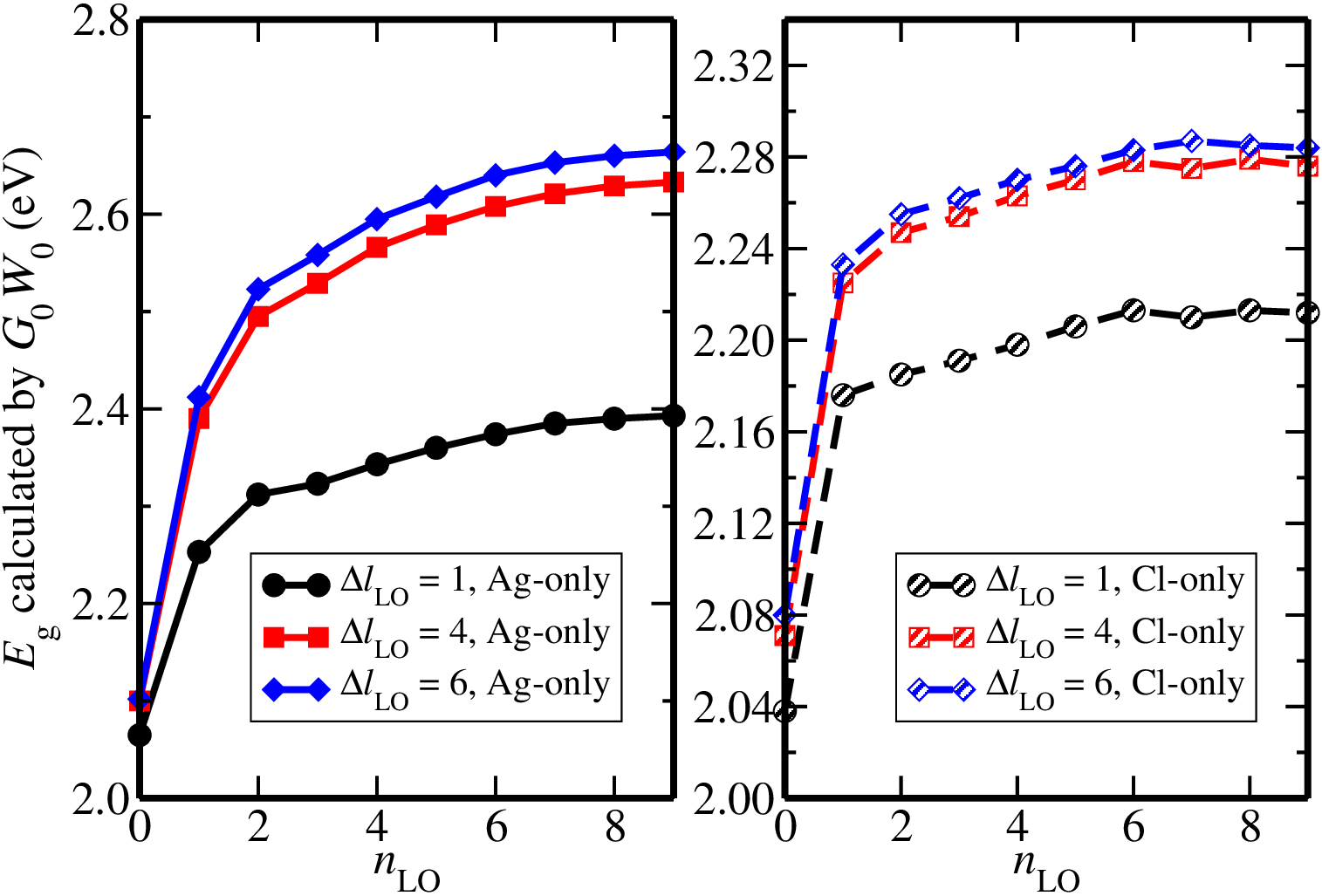}
    \caption{\ Fundamental band gap of AgCl calculated by $G_0W_0$@PBE with HLOs added on either Ag or Cl atoms.}
    \label{figsi:elem-dependence-AgCl-g0w0-gap}  
\end{figure}

\newpage

\begin{figure}[!ht]
    \centering	
    \includegraphics[width=0.9\linewidth]{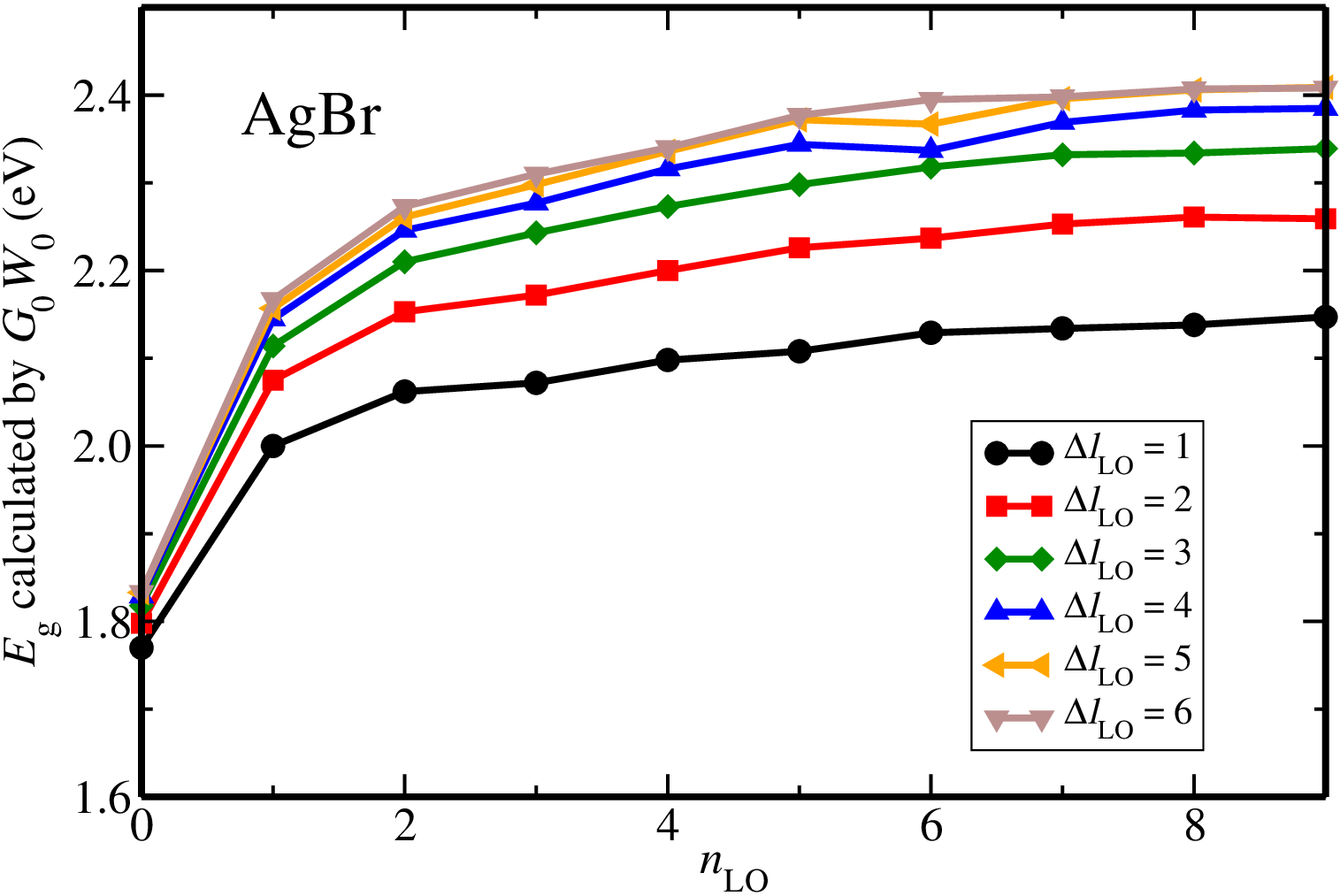}
    \caption{\ Similar to \ref{figsi:CuBr-g0w0-gap-vs-hlo}, but for AgBr.}
    \label{figsi:AgBr-g0w0-gap-vs-hlo}
\end{figure}

\begin{figure}[!ht]
    \centering	
    \includegraphics[width=0.9\linewidth]{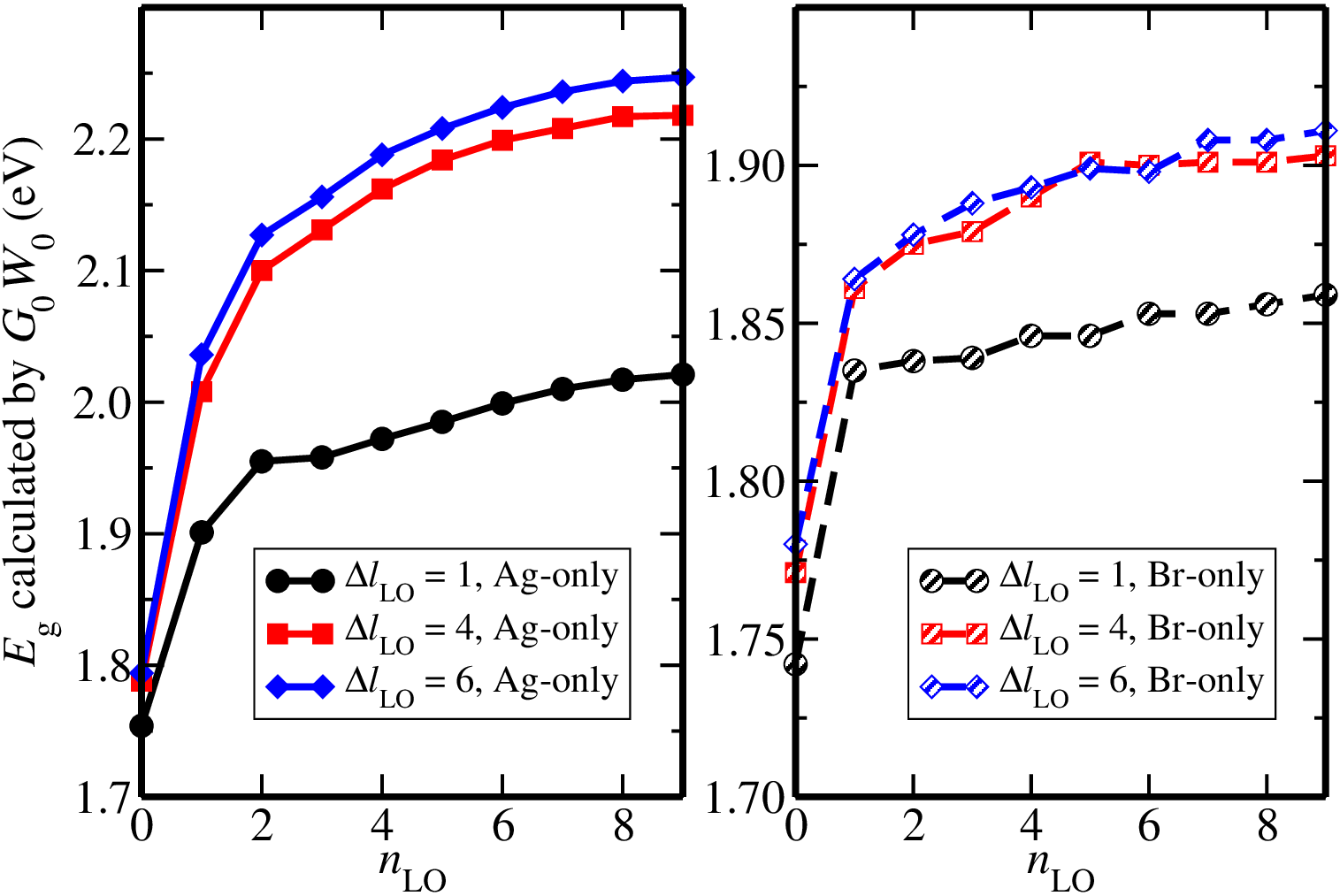}
    \caption{\ Fundamental band gap of AgBr calculated by $G_0W_0$@PBE with HLOs added on either Ag or Br atoms.}
    \label{figsi:elem-dependence-AgBr-g0w0-gap}  
\end{figure}

\newpage

\begin{figure}[!ht]
    \centering	
    \includegraphics[width=0.9\linewidth]{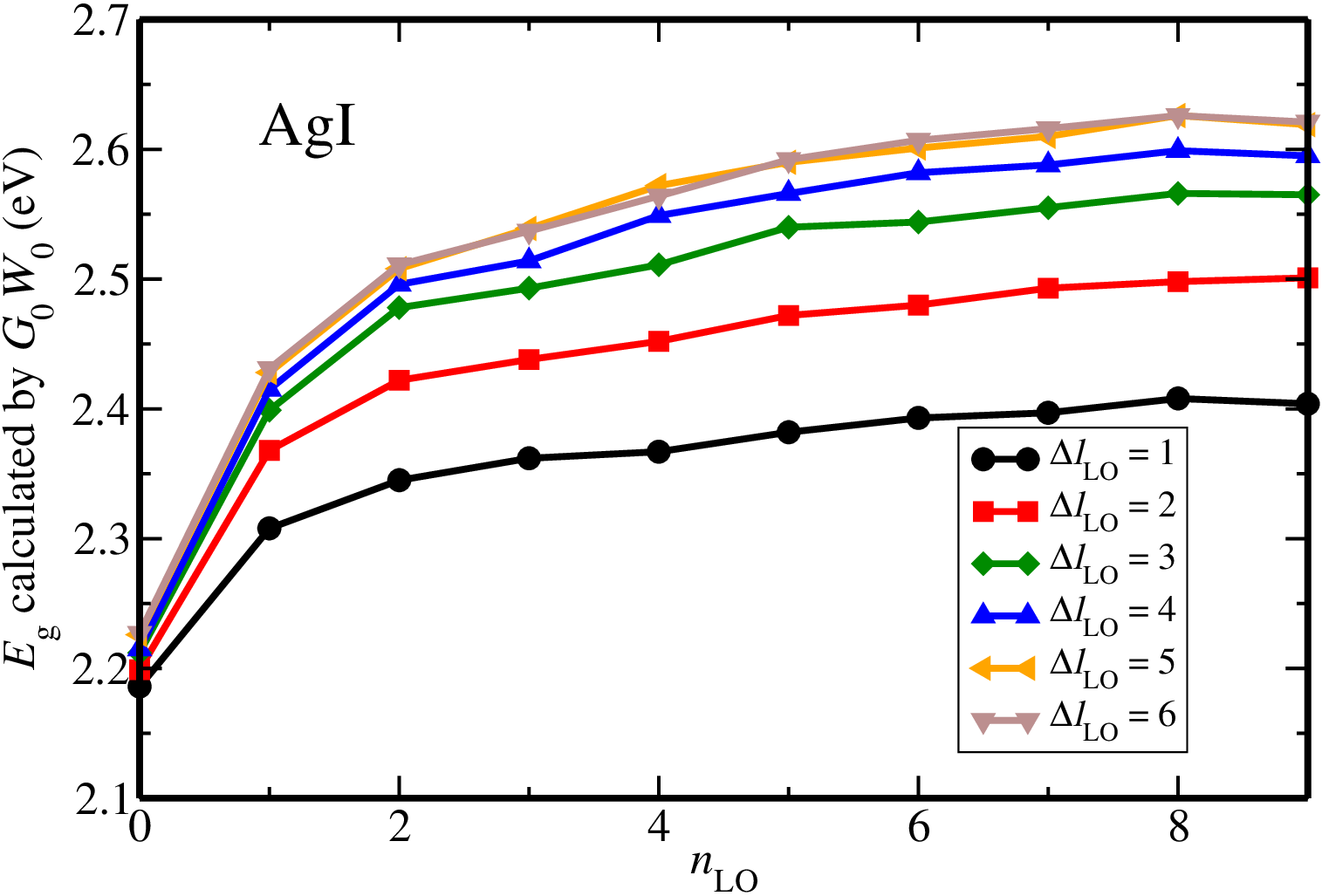}
    \caption{\ Similar to \ref{figsi:CuBr-g0w0-gap-vs-hlo}, but for AgI.}
    \label{figsi:AgIz-g0w0-gap-vs-hlo}
\end{figure}

\begin{figure}[!ht]
    \centering	
    \includegraphics[width=0.9\linewidth]{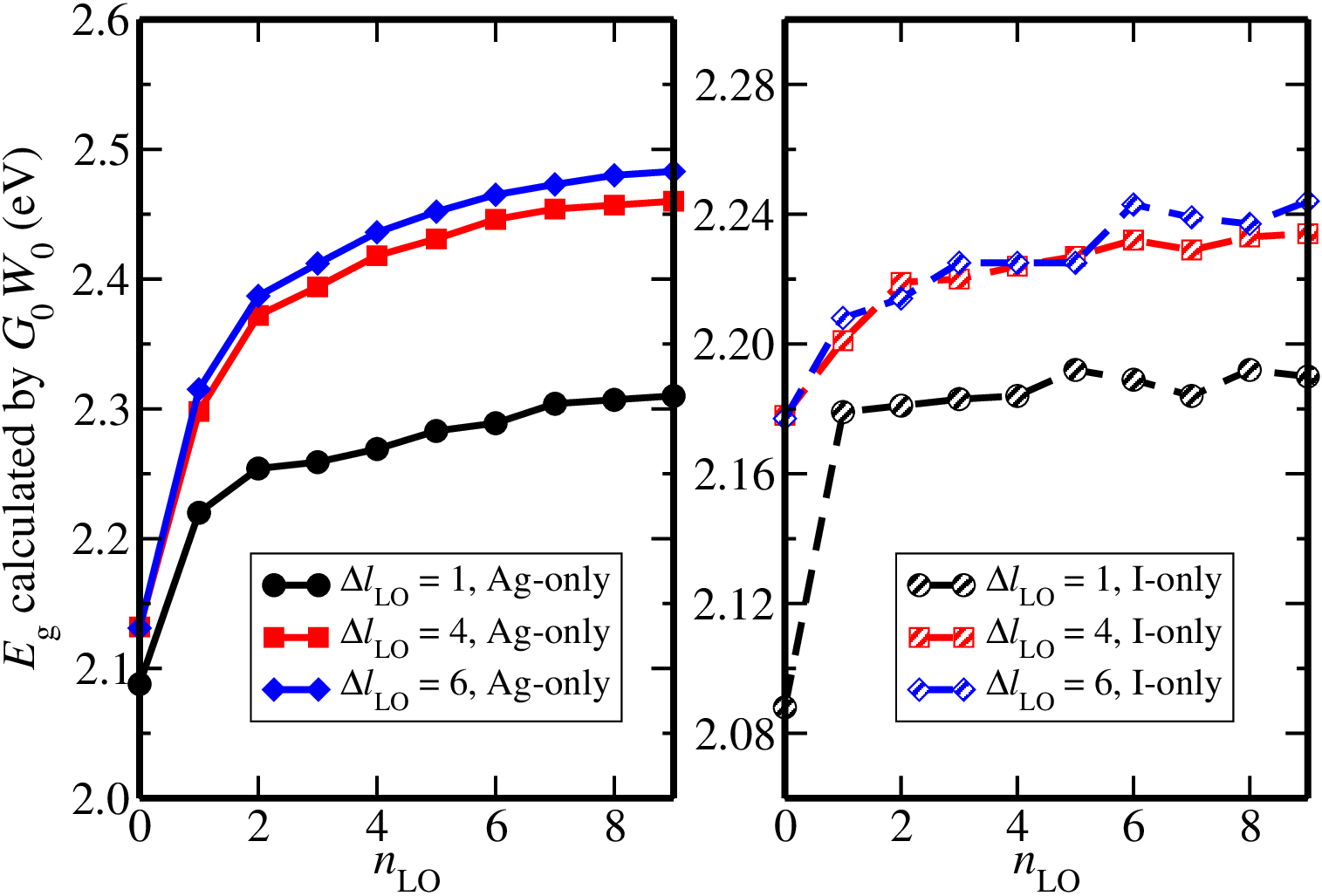}
    \caption{\ Fundamental band gap of AgI calculated by $G_0W_0$@PBE with HLOs added on either Ag or I atoms.}
    \label{figsi:elem-dependence-AgIz-g0w0-gap}  
\end{figure}

\begin{figure}[!ht]
    \centering
    \includegraphics[width=0.9\textwidth]{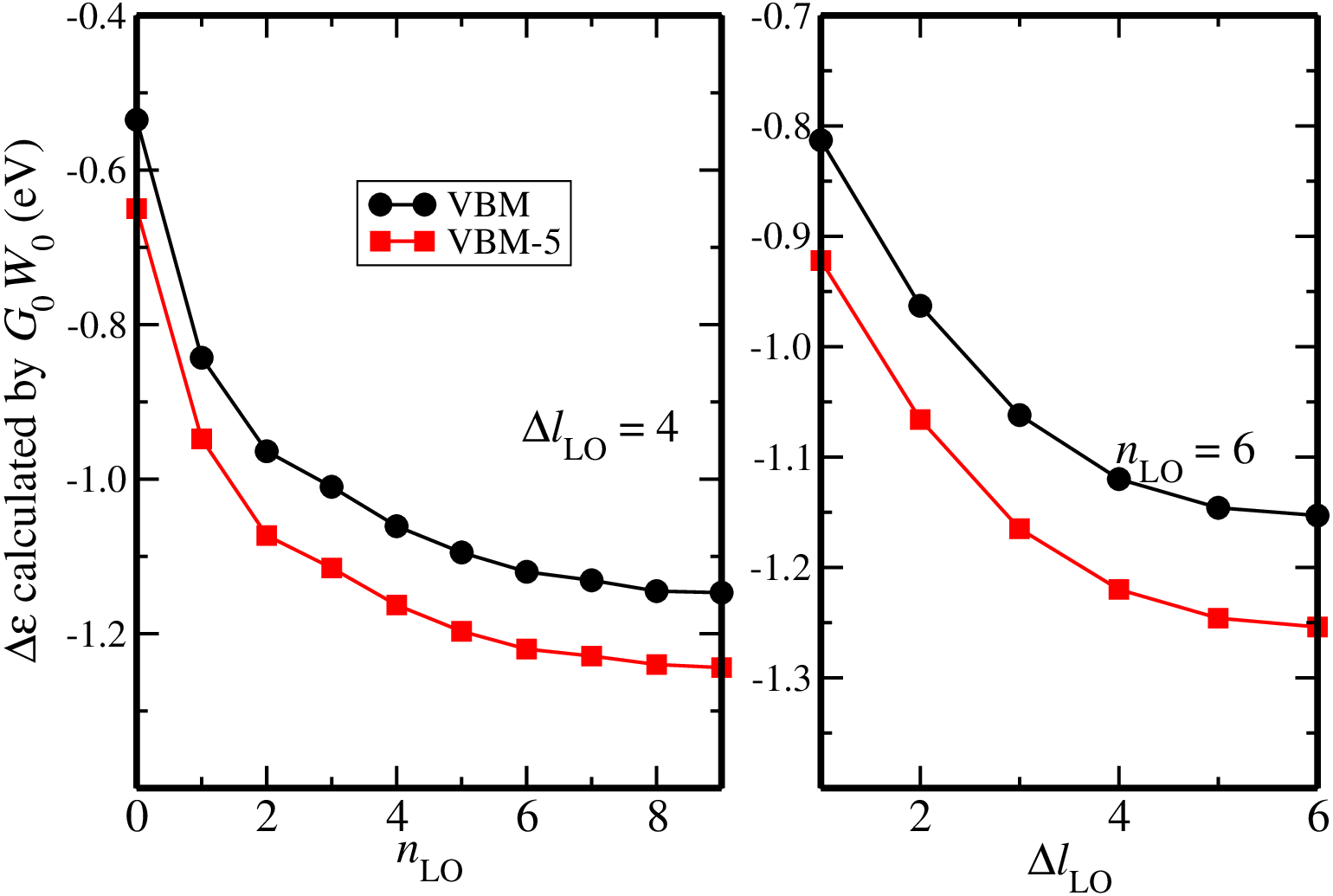}
    \caption{\ Similar to \ref{figsi:CuBr-g0w0-degw-hlo}, but for AgI.}
    \label{figsi:AgIz-g0w0-degw-hlo}
\end{figure}